\documentclass[aps,prx,twocolumn,superscriptaddress,nofootinbib]{revtex4-2}

\usepackage[utf8]{inputenc}
\usepackage[T1]{fontenc}
\usepackage{amsmath,amssymb,amsfonts,mathtools}
\usepackage{bm}
\usepackage{graphicx}
\usepackage{xcolor}
\usepackage{hyperref}
\hypersetup{colorlinks=true,linkcolor=blue,citecolor=blue,urlcolor=blue}

\usepackage{booktabs}
\usepackage{array}

\usepackage{array}

\begin{document}

\title{Testing General Relativity with Individual Supermassive Black Hole Binaries}

\author{Qinyuan Zheng}
\affiliation{Department of Physics, Yale University, New Haven, 06520, CT, USA}
\email{qinyuan.zheng@yale.edu}

\author{Bjorn Larsen}
\affiliation{Department of Physics, Yale University, New Haven, 06520, CT, USA}

\author{Ellis Eisenberg}
\affiliation{Department of Astronomy, Yale University, New Haven, 06520, CT, USA}

\author{Chiara M. F. Mingarelli}
\affiliation{Department of Physics, Yale University, New Haven, 06520, CT, USA}
\affiliation{Center for Computational Astrophysics, Flatiron Institute, 162 5th Avenue, New York, NY, 10010, USA}

\date{\today}

\begin{abstract}
\noindent We develop a unified framework for testing gravity beyond General Relativity (GR) with continuous gravitational waves (CWs) from individual supermassive black hole binaries (SMBHBs). These long-lived, nearly monochromatic nanohertz signals offer unique strengths for precision tests of gravity, since their coherent phase evolution and inter-pulsar correlations in pulsar timing arrays (PTAs) retain detailed information about departures from GR over cosmological propagation distances. We consider three representative classes of deviations from GR: additional polarization states, modified dispersion relations, and parity-violating birefringence. For each, we derive the inter-pulsar cross correlation, the modified antenna response, and the propagation-induced pulsar-term phase delay. For non-tensorial polarizations, the CW cross correlation scales linearly in the alternative-polarization amplitude, compared to the quadratic scaling of the gravitational-wave background (GWB), provided the beyond-GR modes are sub-dominant. PTAs are also competitive for modified dispersion relations, where low frequencies enhance both the antenna-pattern modification and the pulsar-term phase delay. Birefringence, by contrast, is suppressed at nanohertz frequencies for most parity-violating theories. We validate the framework with injection-and-recovery simulations for breathing-mode and massive-graviton signals at current observational limits, recovering the injected beyond-GR parameters and distinguishing the CW signal from both correlated and uncorrelated background models. We further show that a pure-GR CW template recovers source parameters without significant bias when beyond-GR physics is present in the data, supporting a two-stage analysis strategy: identify candidates under GR, then test for deviations.
\end{abstract}

\maketitle

\section{Introduction}
The direct detection of gravitational waves (GWs)~\cite{LIGOScientific:2016aoc} has opened a new window onto fundamental physics, particularly tests of gravity. Observations by ground-based laser interferometers are consistent with General Relativity (GR) to date, placing stringent bounds on many alternative theories of gravity~\cite{LIGOScientific:2019fpa, LIGOScientific:2020tif, LIGOScientific:2021sio}. Nevertheless, GR is generally regarded as an effective description that may break down at sufficiently high energies or large length scales, and a wide range of modified gravity theories has been proposed to address open problems in high-energy physics and cosmology~\cite{Joyce_2016, Capozziello_2011, RevModPhys.82.451}.

Continuous gravitational waves (CWs) from supermassive black hole binaries (SMBHBs) arise in the nanohertz (nHz) frequency band, corresponding to astrophysical length and time scales that are inaccessible to ground-based and most space-based interferometers~\cite{Agazie_2023}. Such signals provide a powerful laboratory for testing gravity~\cite{NANOGrav:2023ygs}. As these massive binaries inspiral slowly over cosmic timescales, they emit long-lived, nearly monochromatic GW signals that can persist for many years to decades. These CW signals, once detected, can be tracked coherently over long observational baselines, enabling precise measurements of phase evolution, polarization response, and propagation effects accumulated over astronomical distances~\cite{Mingarelli2012}. Observational facilities capable of accessing this frequency band---most notably PTAs~\cite{NANOGrav:2023gor, EPTA:2023fyk, Reardon:2023gzh, Xu:2023wog, Miles:2024seg}, and potentially future space-based ultra-low-frequency detectors~\cite{Sesana:2019vho}---therefore offer a unique opportunity to probe gravitational physics using CWs from individual SMBHBs.

In the same frequency range, extensive work has explored beyond-GR searches with the gravitational wave background (GWB)~\cite{Cordes:2024oem, Bernardo:2023zna, Chamberlin:2011ev, Callister_2017, Yunes2025, Inomata:2024kzr, NANOGrav:2023ygs, Liang_2024, Liang:2021bct}. The core idea of these studies is to identify modifications to the Hellings and Downs cross correlation~\cite{1983ApJ...265L..39H} due to deviations from GR. With tentative evidence for a GWB from various PTAs~\cite{NANOGrav:2023gor, EPTA:2023fyk, Reardon:2023gzh, Xu:2023wog, Miles:2024seg}, CWs from individual SMBHBs are a natural next target~\cite{Agazie:2024jbf, Agazie_2023, Agarwal_2026}. This development motivates beyond-GR searches with individual SMBHBs using PTA data~\cite{Liang_2024, Niu_2019, O_Beirne_2019}.

In this work, we derive a new beyond-GR search framework that uses inter-pulsar cross correlation due to an individual SMBHB, a direct extension of~\citet{mingarelli2026fingerprintsindividualsupermassiveblack}. We adopt a unified, theory-agnostic perspective that focuses on observable signatures rather than committing to a specific underlying theory. In particular, we consider three broad physical classes of deviations from GR: (i) alternative GW polarization states beyond the transverse tensor modes of GR~\cite{Takeda_2022, Imafuku_2025, Chamberlin:2011ev, Liang_2024, Niu_2019, O_Beirne_2019, NANOGrav:2023ygs, Chatziioannou:2012rf, Wu_2022}, (ii) modified dispersion relations that alter the frequency dependence of the GW propagation speed and amplitude~\cite{Will_1998, Rubakov_2008, Amelino_Camelia_2001, Magueijo_2002, Amelino_Camelia_2002, Amelino_Camelia_2009, Sefiedgar_2011, Ho_ava_2009_1, Ho_ava_2009_2, VACARU_2012, Blas_2011, Garattini_2011, Garattini_2012, Hendi:2016hbe}, and (iii) chiral GW birefringence arising from parity-violating (PV) modifications of gravity~\cite{Jenks:2023pmk, Ho_ava_2009_1, Ho_ava_2009_2, VACARU_2012, Blas_2011, Nishizawa_2018, Conroy_2019, Alexander:2009tp}, in which the left- and right-handed tensor modes propagate differently. 

By deriving the PTA observables associated with these effects, we clarify how nHz GW observations of individual SMBHBs complement existing constraints from higher-frequency detectors, and we highlight the unique discovery potential of PTAs in the era of CW detections. We find that the multiple projections of the radiation field~\cite{Chatziioannou:2012rf} and ultra-low frequencies of PTAs are advantageous in discerning alternative polarization states and modified dispersion relations. Most notably, for non-tensorial polarizations the CW cross correlation scales linearly in the alternative-polarization amplitude, compared to the quadratic scaling of the GWB. CWs are also able to track phase-evolution and group-velocity deviations induced by modified dispersion relations — capabilities that lie beyond the reach of GWB analyses. On the other hand, frequency-dependent birefringence is highly suppressed in the nHz regime, making higher-frequency space- and ground-based detectors better suited for such tests. Frequency-independent birefringence, however, is worth searching for using PTAs when precise multi-messenger measurements of the SMBHB polarization angle and redshift become available.

Furthermore, we use Bayesian inference to study modified-gravity signals in simulated PTA data, and numerically demonstrate that CW-induced cross correlations can characterize the injected beyond-GR CW signal and distinguish it from both correlated and uncorrelated background models. We also show that the injected beyond-GR parameters can be recovered accurately within our framework. Importantly, our results indicate that it is feasible to first identify the CW sources under the assumption of pure GR, and then test for deviations from GR in a subsequent stage of the analysis.

The paper is structured as follows: in Section~\ref{sec:non-tensor} we derive the cross correlation induced by the non-tensorial polarization modes,  in Section~\ref{sec:dispersion} we discuss the impact of beyond-GR graviton dispersion relations, in Section~\ref{sec:birefringence} we study the inter-pulsar correlation affected by the birefringence due to PV theories, in Section~\ref{sec:injection_and_recovery} we demonstrate the viability numerically by considering two concrete examples: breathing mode and massive gravity. Finally, we discuss and conclude in Section~\ref{sec:discussions}. We use $G=c=1$ unless otherwise stated. Throughout the paper we use $\log_{10}$ for logarithm with base 10, and $\log$ for natural logarithm.

\section{Non-tensorial polarizations}
\label{sec:non-tensor}
In metric theories that permit additional scalar or
vector polarization states, the metric perturbation can be written as~\cite{Eardley1973a, Eardley1973b}
\begin{equation}
h_{ij}(t,\hat{\Omega})
= \sum_P h_P(t)\, e^{P}_{ij}(\hat{\Omega}),
\qquad
P \in \{+,\times,{\rm B},{\rm L},{\rm x},{\rm y}\} ,
\label{eq:general-pol}
\end{equation}
where $\hat{\Omega}$ is the GW propagation direction, and the six modes correspond to the two transverse tensor polarizations $+$ and $\times$, a transverse scalar ``breathing'' mode ${\rm B}$, a longitudinal scalar mode
${\rm L}$, and two transverse vector shear modes ${\rm x}$ and ${\rm y}$ in the
standard classification of metric theories~\cite{Eardley1973a,Eardley1973b,Yunes2025}.
The corresponding timing response follows by integrating the strain over time and summing over the polarization modes:
\begin{equation}
s_a(t) = \sum_P F_a^{P}(\hat{\Omega})\, s_{P, a}(t) ,
\qquad
F_a^{P}(\hat{\Omega})
= \frac{1}{2}
\frac{\hat p_a^i \hat p_a^j}{1+\hat{\Omega}\!\cdot\!\hat{p}_a}\,
e^{P}_{ij}(\hat{\Omega}) .
\end{equation}
Here $F_a^{P}(\hat{\Omega})$ is the antenna pattern function, $\hat{p}_a$ is the unit vector pointing from the Earth to pulsar $a$, and $s_{P, a}(t)$ is the timing residual of mode $P$. Since the pulsar term phase is poorly measured, we focus on the Earth-term cross-correlation in this work~\cite{mingarelli2026fingerprintsindividualsupermassiveblack}. For a circular, non-evolving binary with GW angular frequency $\omega=2\pi f_{\rm gw}$, chirp mass $\mathcal{M}_c$, and luminosity distance $D_L$, the time domain plus and cross strains at the Earth in a polarization basis aligned with 
the binary orbital angular momentum are
\begin{align}
\label{eq:SMBHB waveforms}
h_+(t) &= h_0 (1+\cos^2\iota)\cos(\omega t + \phi_0) ,\\
h_\times(t) &= -2 h_0 \cos\iota \sin(\omega t + \phi_0) ,
\end{align}
where $\iota$ is the inclination angle, $\phi_0$ is an initial phase, and
\begin{equation}
h_0 = \frac{2 \mathcal{M}_c^{5/3}}{D_L} (\pi f_{\rm gw})^{2/3}
\end{equation}
in geometric units. We write down the strains of alternative polarizations at Earth generically as 
\begin{equation}
    h_{(P)}(t)=h_0\mathcal{A}_{(P)}\cos(\omega t + \phi_0 + \delta_{P}),
\end{equation}
where $\mathcal{A}_{(P)}$ is the relative amplitude of the mode normalized to the cross and plus modes, $\delta_{P}$ is the relative phase, and $P$ runs over any beyond-GR modes.  By defining unambiguously 
\begin{equation}
A_{\rm CW} \equiv \frac{h_0}{2\pi f_{\rm gw}}
= \frac{h_0}{\omega}
= \frac{1}{2\pi f_{\rm gw}} \frac{2 \mathcal{M}_c^{5/3}}{D_L} (\pi f_{\rm gw})^{2/3},
\label{eq:A-CW-def}
\end{equation}
we express the Earth term $s_P(t)$ as
\begin{align}
s_P(t) &= A_{\rm CW} \mathcal{A}_{(P)}
\sin(\omega t + \phi_0 + \delta_P).
\end{align}
We note that the full $s_{P, a}(t)$ expression with both the Earth-term and pulsar-term contributions would be pulsar-dependent, while the Earth-term alone has the same phase across the pulsars. Hence we have dropped the subscript $a$.

For a single monochromatic CW the cross correlation generalizes to
\begin{equation}
C_{ab} \equiv \langle s_a s_b\rangle
= \frac{A_{\rm CW}^2}{2}\sum_{P, Q}
\lambda_{ab}^{PQ}F_a^P F_b^Q ,
\label{eq:Cab-general-P}
\end{equation}
where $\lambda_{ab}^{PQ}$ encodes the relative amplitude and phase of each polarization as well as the coupling between different polarization states: $a, b$ run over the pulsars while $P, Q$ run over the polarization modes. For Earth-term only cross correlation, $\lambda_{ab}^{PQ}\rightarrow \lambda^{PQ}$, i.e., it takes the same value for different pulsar pairs.

Although in a full treatment of all contributing polarization modes there are couplings between different polarizations in general, each allowed polarization alone
yields its own antenna pattern and correlation function
$\Upsilon^{(P)}_{ab}$, often with markedly different angular
structure~\cite{Eardley1973a,Eardley1973b,LeeJenetPrice2008,
Chamberlin:2011ev,Wenzer2019,mingarelli2026fingerprintsindividualsupermassiveblack}. In other words, the single-mode correlation functions are the diagonal terms in the polarization space
\begin{equation}
\Upsilon^{(P)}_{ab}=\lambda_{ab}^{PP}F_a^P F_b^P
\end{equation}
Here we \textit{first} derive the correlation functions for non-tensorial polarizations individually, \textit{then} derive the general form for the inter-pulsar cross correlation when multiple beyond-GR polarization modes are present.

For a single polarization state $P$, the Earth–term correlation is simplified to
\begin{equation}
\left\langle s_a(t)s_b(t)\right\rangle
=
\frac{A_{\rm CW}^2}{2}\,
\Upsilon^{(P)}_{ab},
\end{equation}
where the $\lambda_{ab}^{PP}$ factor reduces to a scalar and has been absorbed into the prefactor. The Earth-term cross-correlation pattern is fully determined by
\begin{equation}
\Upsilon^{(P)}_{ab}
\propto
F_a^P\,F_b^P.
\label{eq:Uab-general}
\end{equation}

For a general pulsar position vector $\hat p_a$, we decompose it into longitudinal and transverse components with respect to the propagation direction of the GW $\hat{\Omega}$.
\begin{equation}
\hat p_a=\hat p_{a, \parallel}\, \,+\hat p_{a, \perp}.
\end{equation}
For some transverse basis $\{\hat m,\hat n\}$ corresponding to the GW propagation, the polarization tensors can be represented as
\begin{align}
e^{\rm b}_{ij} &= m_i m_j + n_i n_j, \\
e^{\rm L}_{ij} &= \sqrt{2}\,\Omega_i\Omega_j, \\
e^{\rm x}_{ij} &= m_i\Omega_j + \Omega_i m_j, \\
e^{\rm y}_{ij} &= n_i\Omega_j + \Omega_i n_j.
\end{align}
We may now write down the factor $\hat p_a^{\,i}\hat p_a^{\,j}\,
e^P_{ij}(\hat\Omega)$ for $P=B, L, x, y$. 
For the breathing mode, $\hat p_a^{\,i}\hat p_a^{\,j}\,e^{(B)}_{ij}(\hat\Omega)=(\hat{p}_a\cdot \hat{m})^2+(\hat{p}_a\cdot \hat{n})^2=1-\hat p_{a, \parallel}^2=1-(\hat{\Omega}\cdot \hat{p}_a)^2$. Thus,
\begin{equation}
    F_a^{(B)}(\hat\Omega)=\frac{1}{2}\frac{1-(\hat{\Omega}\cdot \hat{p}_a)^2}{1+(\hat{\Omega}\cdot \hat{p}_a)}.
\end{equation}
For the longitudinal scalar mode, $\hat p_a^{\,i}\hat p_a^{\,j}\,e^{(L)}_{ij}(\hat\Omega)=\sqrt{2}\hat p_{a, \parallel}^2=\sqrt{2}(\hat{\Omega}\cdot \hat{p}_a)^2$, leading to
\begin{equation}
    F_a^{(L)}(\hat\Omega)=\frac{\sqrt{2}}{2}\frac{(\hat{\Omega}\cdot \hat{p}_a)^2}{1+(\hat{\Omega}\cdot \hat{p}_a)}.
\end{equation}
For the vector-x mode, $\hat p_a^{\,i}\hat p_a^{\,j}\,e^{(x)}_{ij}(\hat\Omega)=2\, \hat p_{a, \parallel}\, (\hat p_{a, \perp}\cdot \hat{m})=2\, (\hat{\Omega}\cdot \hat{p}_a)\, (\hat{p}_a\cdot \hat{m})$, giving
\begin{equation}
    F_a^{(x)}(\hat\Omega)=\frac{(\hat{\Omega}\cdot \hat{p}_a)(\hat{p}_a\cdot \hat{m})}{1+(\hat{\Omega}\cdot \hat{p}_a)}.
\end{equation}
Similarly, the vector-y mode has
\begin{equation}
    F_a^{(y)}(\hat\Omega)=\frac{(\hat{\Omega}\cdot \hat{p}_a)(\hat{p}_a\cdot \hat{n})}{1+(\hat{\Omega}\cdot \hat{p}_a)}.
\end{equation}
These expressions are coordinate independent. We note that for the vector modes, the choice of transverse basis affects the form of $\psi$-dependence. If one starts from a basis where $\psi$ happens to be zero, a basis transformation is needed for the final expression (Appendix~\ref{app:basis_transformations}).

\begin{figure}
    \centering
    \includegraphics[width=1\linewidth]{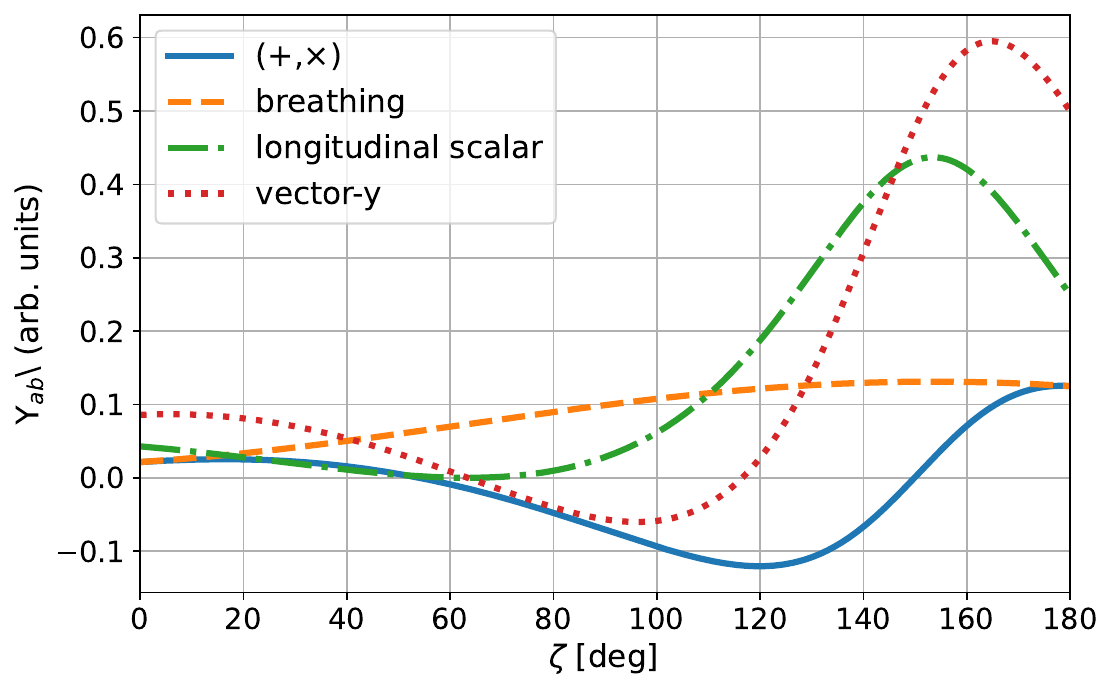}
    \caption{The inter-pulsar correlations induced by the alternative polarizations are distinct from the GR case, offering the basis for detecting beyond-GR polarizations using cross correlations. Here we present the cross-correlation curves for various polarizations in the computational frame, for a source at $(\theta,\phi)=(\pi/4,2\pi/3)$, see Appendix~\ref{app:polarizations}. $F_a^x=0$ in the computational frame by definition, hence the vector x-mode cross correlation is identically zero. }
\label{fig:polarization_modes}
\end{figure}

Fig.~\ref{fig:polarization_modes} illustrates how the non-tensorial components modify the ORF for an example CW, using the computational frame formalism~\cite{Mingarelli_characterizing_2013, Mingarelli14, mingarelli2026fingerprintsindividualsupermassiveblack}. The functions used to compute these may be found in Appendix~\ref{app:polarizations}. Here we have assumed single-mode emission for all the beyond-GR polarizations. The different cross-correlation patterns are the hallmark used to discern the polarization content of a CW.

To account for the inter-mode coupling in the cross correlation pattern, we may start by expressing the timing residual at Earth in pulsar $a$ as
\begin{align}
s_a(t) &= A_{\rm CW} A_a
\sin(\omega t + \phi_0 + \delta_a) + \notag\\
&A_{\rm CW}\sum_{P} \gamma_a^P\sin(\omega t + \phi_0 + \delta_{P})+\notag\\
&A_{\rm CW}\sum_{P} \sigma_a^P\sin\left(\frac{\omega}{2} t + \frac{\phi_0}{2} + \varphi_{P}\right),
\label{eq:sa-A-delta}
\end{align}
where the first line accounts for the tensor mode contribution, the derivation of which can be found in~\cite{mingarelli2026fingerprintsindividualsupermassiveblack}. The relevant parameters are defined as follows:
\begin{align}
A_a &= \sqrt{\alpha_a^2 + \beta_a^2} ,\\
\delta_a &= \arctan\!\left(\frac{\beta_a}{\alpha_a}\right),\\
\alpha_a &= (1+\cos^2\iota)\,F_a^+(\hat{\bm{\Omega}}) ,\\
\beta_a  &= 2\cos\iota\,F_a^\times(\hat{\bm{\Omega}}) ,\\
\gamma_a^P&=\mathcal{A}_{(P)}F_a^{(P)}(\hat{\Omega}) ,\\
\sigma_a^P&=\mathcal{B}_{(P)}F_a^{(P)}(\hat{\Omega})
\end{align}
Here $\mathcal{B}_{(P)}$ is the relative amplitude of the dipolar radiation of a polarization $P$, and $\varphi_P$ is the phase it may acquire with respect to the quadrupolar modes. Note that the antenna patterns here are defined with $\psi=0$, so one needs to transform them accordingly in the appropriate frame, which will introduce $\psi$-dependence in the final expression. Various theories violating the strong equivalence principle and Lorentz invariance include dipolar radiation that is enhanced at ultra-low frequencies~\cite{Will:1977bb, Yunes_2016, Chatziioannou:2012rf}. Hence we also include the dipolar contribution in the timing residual. The cross correlation is then computed by integrating over time and averaging:
\begin{align}
\label{eq:master_correlation}
C_{ab} &= \langle s_a(t)\,s_b(t) \rangle \notag\\
&=\frac{A_{\rm CW}^2}{2}\Big[A_aA_b\cos(\delta_a-\delta_b)\notag\\
&+\sum_P\Big(A_a\gamma_b^P\cos(\delta_a-\delta_{P})\notag\\
&+A_b\gamma_a^P\cos(\delta_b-\delta_{P})\Big)\notag\\
&+\sum_{P,P'}\gamma_a^P\gamma_b^{P'}\cos(\delta_P-\delta_{P'})\notag\\
&+\sum_{P,P'}\sigma_a^P\sigma_b^{P'}\cos(\varphi_P-\varphi_{P'})\Big].
\end{align}
For individual sources, the interference terms of quadrupolar non-tensorial and transverse tensor modes are generally non-vanishing, and are at $\mathcal{O}(\mathcal{A}_P)$. For the GWB correlation, the non-tensorial modes enter at $\mathcal{O}(\mathcal{A}_P^2)$. Hence, individual sources, once detected, generically produce tensor–non-tensor interference terms with favorable scaling, provided the beyond-GR modes are coherent with the tensor modes and subdominant. The dipolar contributions, on the other hand, completely decouple from the quadrupolar terms in the cross correlation.

So far we have focused only on the explicit form of the Earth-term cross-correlation, which is coherent for any pulsar pairs. For a CW source with (approximately) constant GW frequency $f_{\rm gw}$, the
pulsar term enters with a phase shift
\begin{equation}
\Delta\Phi_a 
= 2\pi f_{\rm gw}\,d_a\,\big(1+\hat\Omega\cdot\hat p_a\big),
\label{eq:phase-shift}
\end{equation}
where $d_a$ is the pulsar distance.  

The pulsar term becomes most useful when the pulsar distances are measured with a precision such that the pulsar-term phase uncertainty satisfies~\citep{Lee_2011, BoylePen2012,Mingarelli2012, Taylor:2026}
\begin{equation}
    \sigma(\Delta\Phi_a) \lesssim 1~{\rm rad}.
    \label{eq:phase-criterion}
\end{equation}
The role of the pulsar term differs between stochastic-background searches and the deterministic CW framework developed here.
 For a GWB, the observable correlation is defined only after averaging over an ensemble of sources and integrating over the sky. For tensor polarizations, this sky average causes the pulsar term to drop out of the overlap reduction function, yielding the familiar Hellings–Downs form. This mechanism, however, is not generic: for certain alternative polarizations, particularly the longitudinal mode,
\begin{equation}
F^{\rm L}_a(\hat\Omega)
\propto 
\frac{\big(\hat\Omega\cdot\hat p_a\big)^2}{1+\hat\Omega\cdot\hat p_a},
\label{eq:longitudinal-behavior}
\end{equation}
the sky integral is sensitive to singular source directions, and the pulsar term must be retained in order for the overlap reduction function to remain well defined, as discussed in~\cite{Chamberlin:2011ev}. In the CW case, by contrast, no such sky average is performed, since the source is modeled as an individual SMBHB at a definite sky location. The issue is therefore not one of singular behavior in a sky-integrated response. Rather, the practical reason for omitting the pulsar term is that its phase is typically poorly constrained by present pulsar-distance uncertainties, making it appropriate to marginalize over as an effectively unknown contribution~\cite{corbin2010pulsartimingarrayobservations, Mingarelli2012, Lee_2011, charisi2023efficientlargescaletargetedgravitationalwave}. The justification is therefore rooted in phase uncertainty, not in the detailed antenna response of a given polarization mode. Therefore, as long as the CW propagation direction does not coincide with the direction to the pulsar, which would induce a singularity in the antenna pattern, the Earth-term cross-correlation remains a useful and well-defined observable when pulsar term measurements are unavailable.

\section{Dispersion relation}
\label{sec:dispersion}
In GR, the graviton is massless and follows the dispersion relation
\begin{equation}
\omega^2 = k^2c^2.
\end{equation}
Throughout this section we have set $\hbar=1$ and kept $c$ explicit for clarity of the velocity modification. In certain beyond-GR gravities, including some parity-violating theories considered in the previous section, the dispersion relation would take a different form~\cite{Will_1998, Rubakov_2008, Amelino_Camelia_2001, Magueijo_2002, Amelino_Camelia_2002, Amelino_Camelia_2009, Sefiedgar_2011, Ho_ava_2009_1, Ho_ava_2009_2, VACARU_2012, Blas_2011, Garattini_2011, Garattini_2012, Hendi:2016hbe}. One may take a general parameterization of the dispersion relation and map it to various modified gravity theories, while a modified dispersion relation alone leaves an imprint on the cross correlation. A generic form of the modified graviton dispersion relation is~\cite{Mirshekari_2012}
\begin{equation}
\label{eq: dispersion_relation}
\omega^2 = k^2c^2+m_g^2 c^4+\mathbb{A}k^\alpha c^\alpha,
\end{equation}
where $m_g$ is the mass of the graviton, $\mathbb{A}$ and $\alpha$ are Lorentz-violating parameters that characterize the deviation from GR. We assume that the deviation is small, i.e., $\mathbb{A}/(ck)^{2-\alpha}\ll1$. 

Different beyond-GR theories assign different values to $m_g$, $\mathbb{A}$, and $\alpha$. The most intuitive modification is a massive graviton, such that 
\begin{equation}
\omega^2 = k^2c^2+m_g^2 c^4.
\end{equation}
For other examples of how various theories' dispersion relations can be cast into this form, see~\cite{Mirshekari_2012}.

With the generic parameterization of the dispersion relation, the wave vector is a function of the angular frequency of the GW $\bm{k}=\bm{k}(\omega)$. Starting from the time component of the null geodesic equation for photons~\cite{Hobbs_2009, Lee_2010},
\begin{equation}
\frac{d\omega_p}{d\lambda}
+ \frac{\omega_p^2}{2}
\frac{\partial h_{ij}}{\partial t}
\hat p^i \hat p^j
=0 ,
\end{equation}
where $\omega_p$ is the photon angular frequency, $\hat p^i$ is the unit vector
pointing from the observer to the pulsar, and $\lambda$ is an affine parameter
normalized such that $dt/d\lambda=\omega_p$ in the Minkowski background.

\begin{figure*}
    \centering
    \includegraphics[width=0.49\linewidth]{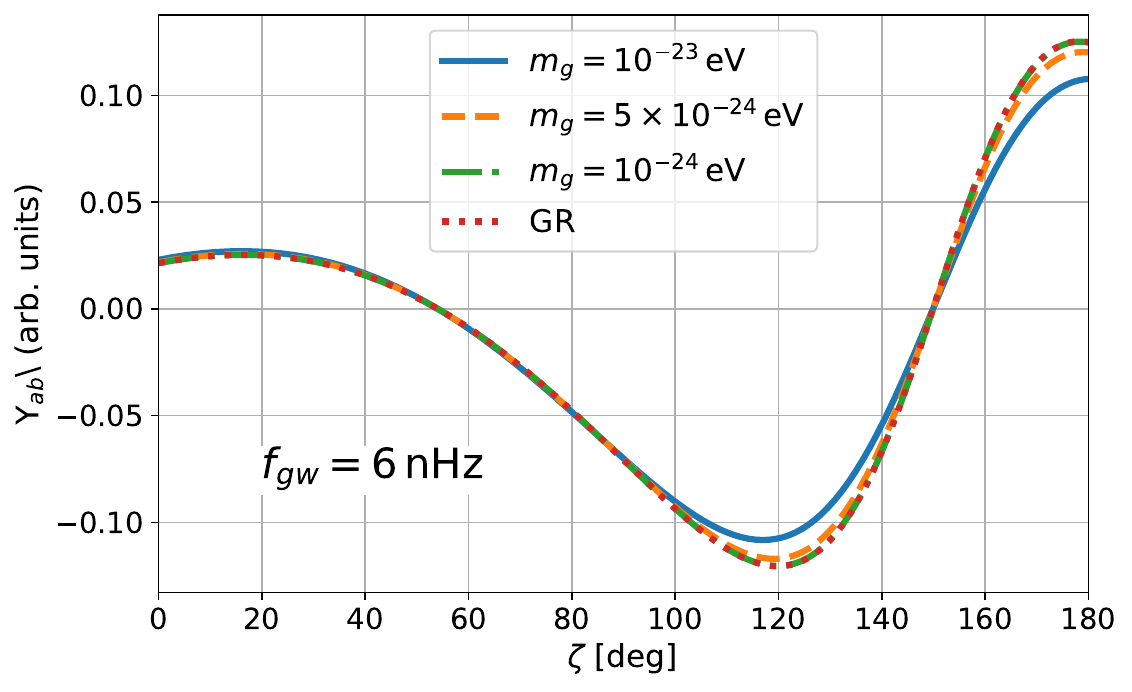} \includegraphics[width=0.49\linewidth]{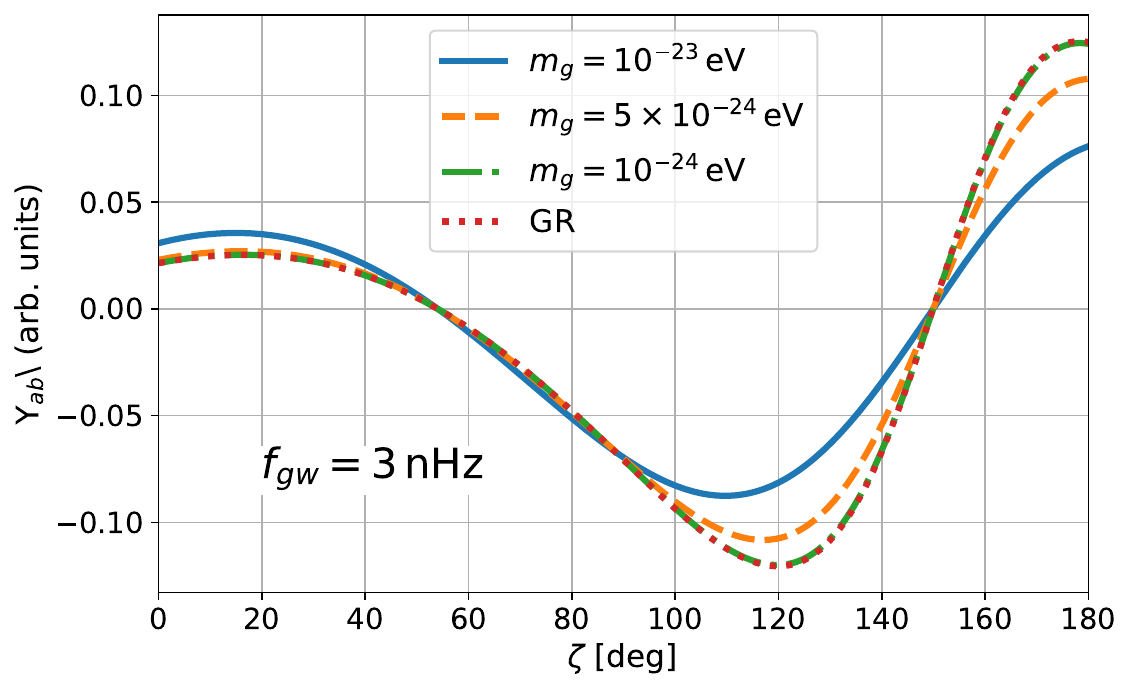}
    \caption{The inter-pulsar correlations induced by massive gravity are distinct from GR, and GW frequency dependent. The lower the frequency is, the larger the deviation from GR is. This frequency dependence is generally true for modified dispersion relations. Here we present the cross-correlation curves for various graviton masses in the computational frame, for a source at $(\theta,\phi)=(\pi/4,2\pi/3)$.}
\label{fig:massive_gravity_cross_correlation}
\end{figure*}

To leading order in the metric perturbation, the photon trajectory can be
approximated by the unperturbed path
\begin{equation}
\mathbf{x}(t)= - \hat{p}\, c t .
\end{equation}
Integrating the geodesic equation along the photon trajectory from the pulsar
(at distance $|\mathbf d_a|$) to the observer gives the fractional frequency shift
\begin{equation}
\frac{\Delta \omega_p(t)}{\omega_p}
= \frac{1}{2}\hat p^i \hat p^j
\int_{t-|\mathbf d_a|/c}^{t}
\partial_t h_{ij}(t',\mathbf{x}(t')) dt'.
\end{equation}

For a monochromatic plane gravitational wave
\begin{equation}
h_{ij}(t,\mathbf{x}) =
h_{ij} e^{i(\omega t-\mathbf{k}\cdot \mathbf{x})},
\end{equation}
evaluating the integral along the photon path introduces the factor
\begin{equation}
1+\frac{\bm{k}(\omega)c}{\omega}\cdot \hat{p}
\end{equation}
The resulting frequency shift becomes
\begin{equation}
\frac{\Delta\omega_p(t)}{\omega_p}
=
\frac{1}{2}\frac{\hat{p}^i\hat{p}^j}{1+\frac{\bm{k}(\omega)c}{\omega}\cdot \hat{p}}
\left[
h_{ij}(t,0)
-
h_{ij}\!\left(t-\frac{|\bm{d_a}|}{c},\bm{d_a}\right)
\right],
\end{equation}

The timing residual may be recast as 
\begin{equation}
    s_a(t)=\frac{1}{2}\frac{\hat{p}_a^i\hat{p}_a^j}{1+\frac{\bm{k}(\omega)c}{\omega}\cdot \hat{p}_a}\left[s_{ij}(t,0)-s_{ij}\left(t-\frac{|\bm{d_a}|}{c},\bm{d_a}\right)\right]
\end{equation}
with
\begin{equation}
    s_{ij}(t)=\sum_{P}\int^t d\tau \, h^P_{ij}(\tau)=\sum_P s^P(t) e_{ij}^P
\end{equation}
where $P$ iterates over all the polarizations. A modified Hellings-Downs correlation for the GWB with massive gravity was first derived by~\citet{Lee_2010}. Here we derive the CW cross correlation with a generic modified dispersion relation. The Earth term cross correlation can be written in the same form as that in Sec.~\ref{sec:non-tensor}, and reads 
\begin{align}
\label{eq:dispersion_correlation}
C_{ab} &= \langle s_a(t)\,s_b(t) \rangle \notag\\
&=\frac{A_{\rm CW}^2}{2}\Big[A_aA_b\cos(\delta_a-\delta_b)\notag\\
&+\sum_P\Big(A_a\gamma_b^P\cos(\delta_a-\delta_{P})\notag\\
&+A_b\gamma_a^P\cos(\delta_b-\delta_{P})\Big)\notag\\
&+\sum_{P,P'}\gamma_a^P\gamma_b^{P'}\cos(\delta_P-\delta_{P'})\notag\\
&+\sum_{P,P'}\sigma_a^P\sigma_b^{P'}\cos(\varphi_P-\varphi_{P'})\Big],
\end{align}
where the parameters have the same definition, except for the antenna pattern functions, for which we now have
\begin{align}
    F_a^{P}(\hat{\bm{\Omega}})
&= \frac{1}{2}
\frac{\hat p_a^i \hat p_a^j}{1+\frac{\bm{k}(\omega)c}{\omega}\!\cdot\!\hat{p}_a}\,
e^{P}_{ij}(\hat{\bm{\Omega}})\notag\\
&=\frac{1}{2}
\frac{\hat p_a^i \hat p_a^j}{1+\frac{c}{v_p(\omega)}\hat \Omega\!\cdot\!\hat{p}_a}\,
e^{P}_{ij}(\hat{\bm{\Omega}})
\end{align}
where $v_p$ is the phase velocity of the GW. In Fig.~\ref{fig:massive_gravity_cross_correlation} we present the modification to the cross correlation induced by a massive graviton. The larger the graviton mass is, the more the correlation curve deviates from GR. We also note that the deviations due to the modified dispersion, such as in massive gravity, are generally GW-frequency-dependent. In the example presented here, the lower the GW frequency is, the larger the modification to the inter-pulsar cross correlation.

The modified dispersion relation will further alter the relative phase between the Earth term and the pulsar term. For a monochromatic signal of frequency $f_{\rm gw}$, the Earth-term phase remains 
$2\pi f_{\rm gw}t=\omega t$, while the pulsar term becomes
\begin{equation}
\Phi_{a,{\rm P}}
= \omega\!\left(t - \frac{d_a}{c}\,\left(1+\frac{\bm{k}(\omega)c}{\omega}\!\cdot\!\hat{\bm{p}}_a\right)\right),
\end{equation}
introducing an additional delay
\begin{align}
\label{eq:dispersion_phase}
\Delta\Phi^{\rm Disp}_a
&= -\frac{d_a}{c}\left(c\bm{k}(\omega)-\omega\hat{\bm{\Omega}}\right)\cdot\hat{p}_a,
\end{align}
This additional phase delay in the pulsar term depends on both the frequency and the pulsar-Earth geometry. For some choices of $m_g, \mathbb{A}$ and $\alpha$, the wave vector $k$ becomes imaginary, indicating that the wave attenuates and does not propagate.

Finally, the modified graviton group velocity to the first order in $\mathbb{A}$ is
\begin{equation}
    \frac{v_g^2}{c^2}\simeq1-\frac{m_g^2c^4}{\omega^2}+(\alpha-1)\mathbb{A}\omega^{\alpha-2}\Big(1-\frac{m_g^2c^4}{\omega^2}\Big)^{\alpha/2}.
\end{equation}
We note that in addition to the group velocity, the literature has sometimes used the ``particle velocity'' $v_{\rm prtl}=c^2/v_p$ to describe the propagation of different frequency modes. This convention is equivalent to the group velocity if only the mass correction term is present, and fails the wavepacket propagation interpretation once the Lorentz-violating term in Eq.~(\ref{eq: dispersion_relation}) is nonzero. For a detailed discussion, see~\cite{Ezquiaga:2022nak}.

The correction to the group velocity could be especially important for multi-messenger observations of GW events, as the arrival times of the EM and GW signals of the same events will differ. Making inferences from phase-shift measurements requires a precise measurement of the pulsar term, which in turn necessitates precise measurements of the pulsar distances. Despite these technical requirements, ultra-low-frequency CWs are advantageous for searching for certain modified dispersion relations. For example, for the massive gravity, the correction to the group velocity of GWs is suppressed by $f_{\rm gw}^2$. For typical PTA observations at $10\, \rm nHz$ and ground-based detectors operating at $100\, \rm Hz$, this scaling implies a $10^{20}$-fold difference between the massive-gravity-induced modifications of the antenna pattern in the PTA band and ground-based detectors.

\section{Birefringence}
\label{sec:birefringence}
In GR, GW polarization states are preserved as the wave propagates through a homogeneous, isotropic background. If gravity is coupled to a field that breaks chiral symmetry, however, the left- and right-handed polarizations evolve differently during propagation, in both relative phase and amplitude~\cite{Zhao_2020, Jenks:2023pmk}. These propagation-induced differences between the chiral states are termed birefringence: changes in the relative phase are called velocity birefringence, and changes in the relative amplitude are called amplitude birefringence.

In the form of an effective field theory (EFT), the modification of GR can be generically summarized as 
\begin{equation}
    S = \frac{1}{16\pi G}\int d^4x \sqrt{-g} \Big[\mathcal{L}_{GR}+\mathcal{L}_{PV}+\mathcal{L}_{other}\Big],
\end{equation}
where $\mathcal{L}_{GR}$ is the Einstein-Hilbert term, $\mathcal{L}_{PV}$ are the terms that break the parity symmetry, and $\mathcal{L}_{other}$ contains all the other terms such as the matter fields and modified gravity terms irrelevant to parity violation.

For instance, Chern-Simons (CS) gravity~\cite{Alexander:2009tp, Lue_1999, PhysRevD.68.104012} is the leading order effective extension of GR with a parity-violating term
\begin{equation}
\label{eq:CS_term}
    \mathcal{L}_{PV}^{CS}=\frac{1}{8}a(\phi)\varepsilon^{\mu\nu\rho\sigma}R_{\rho\sigma\alpha\beta}R^{\alpha\beta\mu\nu}.
\end{equation}
Here $\varepsilon^{\mu\nu\rho\sigma}$ is the Levi-Civita antisymmetric tensor, $R$ is the Riemann tensor, and $a(\phi)$ is an arbitrary function of a scalar field $\phi$, which in turn is a function of spacetime. Note that when $a(\phi)=\mathrm{const}$, the CS theory reduces to GR since the PV term can be eliminated by field redefinition and integration by parts.

\begin{figure*}
    \centering
    \includegraphics[width=0.9\linewidth]{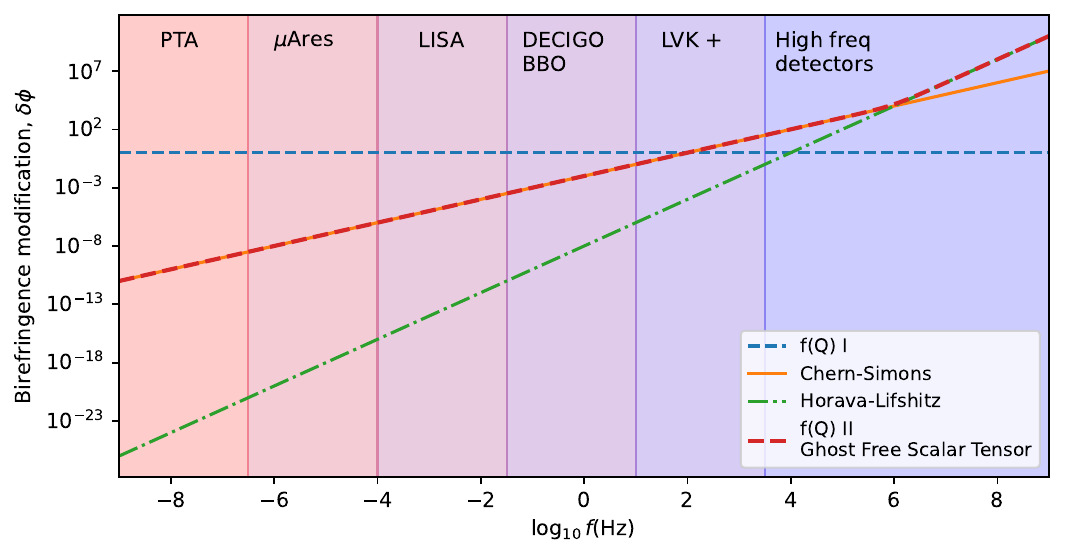}
    \caption{Most birefringent theories are suppressed at ultra-low frequencies, making high-frequency detectors better detection apparatus~\cite{Zhao_2022, LIGOScientific:2021sio}. The frequency-independent birefringence (e.g. f(Q) I theory) is fully degenerate with the source polarization angle, requiring further work to break the degeneracy. Here we plot the birefringence-induced modification to the GR waveform set by the current limits on various beyond GR theories~\cite{Jenks:2023pmk, Zhao_2022, PhysRevD.108.084068, Wang:2025fhw}, as a function of GW frequency. The birefringent modification is stronger for higher redshift sources, and we take a fiducial redshift of $z=1$ for the CW source here (Eq.~\ref{eq:PV_parameterization}). }
\label{fig:birefringence}
\end{figure*}

Other alternative theories of gravity---such as Horava-Lifshitz gravity~\cite{Ho_ava_2009_1, Ho_ava_2009_2, VACARU_2012, Blas_2011}, ghost-free scalar-tensor gravities~\cite{Nishizawa_2018}, symmetric teleparallel equivalence of GR theory~\cite{Conroy_2019}---can similarly be cast into this form and give rise to birefringence.

In this study, we focus on a generic treatment of parity-violating theories and derive how birefringence of GWs from an individual SMBHB is imprinted on the inter-pulsar cross correlation. In PV theories, the polarizations received at the detector are related to the polarizations at emission by~\cite{Jenks:2023pmk}
\begin{align}
    h_R(\omega)&=h_R^{GR}(\omega)e^{-\delta\phi_A(\omega)-i\delta\phi_V(\omega)}\notag\\
    h_L(\omega)&=h_L^{GR}(\omega)e^{\delta\phi_A(\omega)+i\delta\phi_V(\omega)}
\end{align}
where we assume that the source effects due to the modified gravity are small so that the frequency-domain circular polarizations $h_{R, L}^{GR}(\omega)$ are the same as GR. Here $\delta\phi_V(\omega)$ captures the relative phase shifts between the two polarizations---the velocity birefringence, while $\delta\phi_A(\omega)$ encodes the relative amplitude shifts---the amplitude birefringence.

Assuming a circular orbit, the response of a pulsar can be written as 
\begin{align}
\label{eq:h_af}
h_a(\omega)
= h_a^{\mathrm{GR}}
\left[
1
+ f(F_a^{+, \times}, \xi)\,\delta\phi_{A}(\omega)
- g(F_a^{+, \times}, \xi)\,\delta\phi_{V}(\omega)
\right]\notag\\
\exp\!\left\{
i \left[
g(F_a^{+, \times}, \xi)\,\delta\phi_{A}(\omega)
+ f(F_a^{+, \times}, \xi)\,\delta\phi_{V}(\omega)
\right]
\right\}
\end{align}
where
\begin{align}
f(F_{+, \times}, \xi)
&= \frac{2\left(F_{+}^{2}+F_{\times}^{2}\right)\,\xi\left(1+\xi^{2}\right)}
{4F_{\times}^{2}\xi^{2}+F_{+}^{2}\left(1+\xi^{2}\right)^{2}}, \\
g(F_{+, \times}, \xi)
&= \frac{F_{+}F_{\times}\left(-1+\xi^{2}\right)^{2}}
{4F_{\times}^{2}\xi^{2}+F_{+}^{2}\left(1+\xi^{2}\right)^{2}} .
\end{align}
Here $\xi=\cos\iota$, and $h_a^{GR}(\omega)$ is the GR response, given by

\begin{equation}
\label{eq:GR_decomp}
    h_a^{GR}=F_a^{+}h_{+}\, +F_a^{\times}h_{\times}
\end{equation}

The timing residual is given by integrating the strain over time
\begin{equation}
\label{eq:s_at}
    s_a(t)=\int^t dt'\int d\omega \, h_a(\omega) e^{i\omega t'}.
\end{equation}

\begin{table}[t]
\centering
\begin{tabular}{|l|c|c|c|c|c|}
\hline
\textbf{Theory} & $\alpha$ & $\beta$ & $\rho$ & $\sigma$ \\
\hline

Chern-Simons~\cite{Alexander:2009tp, Lue_1999, PhysRevD.68.104012}&
\checkmark &
0 & 0 & 0\\

Ghost-Free Scalar Tensor~\cite{Nishizawa_2018} &
\checkmark &
\checkmark &
0 &
\checkmark\\

Symmetric Teleparallel I~\cite{Conroy_2019}&
0 & 0 &
\checkmark &
0\\

Symmetric Teleparallel II~\cite{Conroy_2019} &
\checkmark &
\checkmark &
0 &
\checkmark\\

Horava-Lifshitz~\cite{Ho_ava_2009_1, Ho_ava_2009_2, VACARU_2012, Blas_2011}&
0 & 0 & 0 &
\checkmark \\

\hline
\end{tabular}
\caption{Parameterizations of various parity-violating modified gravity theories, to the leading order beyond GR as a parametrized expansion power counted by the PV energy scale $\Lambda_{\rm PV}$~\cite{Jenks:2023pmk, Zhao_2020}. \checkmark indicates a non-zero contribution.}
\label{table: PV theories}
\end{table}

Given Eq.~(\ref{eq:Cab-general-P}) for the cross correlation, we may derive the general expression of the cross correlation with birefringences. As a low energy effective theory, $\delta\phi_{V, A}$ may be organized as a parametrized expansion in terms of the cut-off energy scale~\cite{Jenks:2023pmk}. As an approximation, we retain only the leading order terms and obtain
\begin{align}
\label{eq:PV_parameterization}
\delta \phi_A
&= k(1+z)\Big(\alpha z+\frac{\beta d_2(z)}{1+z}\Big), \notag\\
\delta \phi_V
&= \rho \ln(1+z)+\sigma \,k^2(1+z)^2d_3(z).
\end{align}

Here $k$ is the wave number, $\alpha$, $\beta$, $\rho$, $\sigma$ are coefficients of the parametrized expansion that can be matched to the parameters of specific gravity theories, and $z$ is the redshift of the source. Here we have retained terms up to the order $\Lambda_{PV}^{-2}$ and absorbed the PV energy scale $\Lambda_{PV}$ into the Wilson coefficients from~\cite{Jenks:2023pmk}. $d_2(z)$ is the angular-diameter distance at redshift $z$ and $d_3(z)=(1+z)^{-2}\int (1+z)^{1}H^{-1}(z) dz$. To the leading order, $\delta\phi_A$ is linear in $k$ and $\delta\phi_V$ is a constant with respect to $k$. We note that for certain modified gravity theories the form can be further simplified. For example, for CS gravity, ghost free scalar-tensor theory, symmetric teleparallel theories and Horava-Lifshitz theory, $\sigma=0$. For CS gravity, the only non-vanishing parameter is $\alpha$. In Table~\ref{table: PV theories}, we list the parameterizations of a few frequently considered PV theories.

Plugging Eq.~(\ref{eq:h_af}) into Eq.~(\ref{eq:s_at}), we compute the Earth term cross correlation with birefringence (see Appendix~\ref{app:birefringence} for details), which takes the form:
\begin{equation}
    C_{ab}=\langle s_a(t)s_b(t) \rangle_T=\frac{A^a_{CW}A^b_{CW}}{2}\cos(\delta_a-\delta_b).
\end{equation}
Unlike in GR, $A^{a, b}_{CW}$ and $\delta_{a, b}$ are in general frequency dependent. They also depend on the effective parameters $\alpha, \beta, \rho, \sigma$, which in turn depend on the specific gravity theory under consideration. In principle, the framework presented here accounts for the leading order contributions in the parametrized expansion in terms of PV energy scale and is theory-agnostic, amounting to a robust search for birefringence, regardless of which parity-violating theory of gravity is at work. In addition to the Earth-term cross correlation, the modified dispersion relation will also modify the antenna pattern as well as introduce a phase shift in the pulsar term. However, such effects are tiny compared to the accumulated propagation effects given the existing limits on the expansion coefficients, which we show in detail in Appendix~\ref{app:birefringence}.

As one of the most studied theories of modified gravity, CS gravity is a leading-order correction to GR that includes a gravitational PV term (see Eq.~(\ref{eq:CS_term})) and is considered the ``vanilla'' version of a PV theory~\cite{Lue_1999, PhysRevD.68.104012, Alexander:2009tp}. In the low-energy parametrized framework we adopt, the CS correction is parameterized by a non-zero $\alpha$ to the leading order. The most stringent limit on the CS parity violation comes from analyzing the GWTC-3 data, which results in a measurement of the birefringence opacity parameter $\kappa = -0.019^{+0.038}_{-0.029} \,\mathrm{Gpc}^{-1}$ at $100\, \rm Hz$~\cite{PhysRevD.108.084068}. This limit translates to an upper limit of $\alpha\lesssim\mathcal{O}(10^{-2})\,\rm Hz^{-1}$, which we take as the benchmark value for an order-of-magnitude estimate.

For CS modified gravity, the dispersion relation is unchanged~\cite{Jenks:2023pmk}, and Eq.~(\ref{eq:varphidef}) reduces to
\begin{align}
\label{eq:CS_varphi}
    \varphi_a(\omega,\vec{\theta})= \left[
1
+ \alpha\, f(F_a^{+, \times}, \xi)\,\omega(1+z)z
\right]\notag\\
\exp\!\left\{
i \alpha\, 
g(F_a^{+, \times}, \xi)\,\omega(1+z)z
\right\},
\end{align}
which gives
\begin{align}
\label{eq: p_and_q}
    &p_a =\notag\\
     &\left[1+ \alpha\, f(F_a^{+, \times}, \xi)\,\omega(1+z)z\right]\cos\left[\alpha\, 
g(F_a^{+, \times}, \xi)\,\omega(1+z)z\right],\notag\\
    &q_a = \notag\\
    &\left[1+ \alpha\, f(F_a^{+, \times}, \xi)\,\omega(1+z)z\right]\sin\left[\alpha\, 
g(F_a^{+, \times}, \xi)\,\omega(1+z)z\right].
\end{align}
Plugging this into the Eq.~(\ref{eq:delta_and_A}), we find the CS induced correction to GR is highly suppressed in the PTA band due to the linear scaling with the frequency: with $\alpha\lesssim\mathcal{O}(10^{-2})\,\rm Hz^{-1}$ at $10^{-8}\, \rm Hz$, the CS correction is suppressed by a factor of $\mathcal{O}(10^{-10})$, which practically renders the signal undetectable by PTAs, assuming that the theoretical model and parameterization used here are an accurate infrared approximation. 

For the frequency-dependent PV corrections considered above, the birefringent modification is strongly suppressed in the PTA band. Starting with Eq.~(\ref{eq:PV_parameterization}), we define the following quantity to characterize the amplitude of birefringence induced modification to the observed waveforms:
\begin{equation}
    \delta\phi = \sqrt{\delta\phi_A^2+\delta\phi_V^2}.
    \label{eq:deltaphi}
\end{equation}
Fig.~\ref{fig:birefringence} shows the frequency dependence of  $\delta\phi$ for various PV modified gravity theories, where we have used current limits on the expansion coefficients derived from constraints in~\cite{Jenks:2023pmk, Zhao_2022, PhysRevD.108.084068, Wang:2025fhw}. It clarifies that the low-frequency suppression of birefringence modifications is not limited to the CW cross correlation method presented above, but applies to low-frequency observations in general. The implications are twofold. On the one hand, the frequency suppression makes ultra-low-frequency observations like PTAs ineffective in the search for birefringence induced by PV theories, compared to ground-based interferometers; on the other hand, it eliminates a significant class of confusion signals that we would have to consider when searching for other beyond-GR signals, reducing the search degeneracies in the theory space.

This low-frequency suppression is absent if the birefringence becomes frequency-independent. One such example is the symmetric teleparallel I (f(Q) I) theory~\cite{Conroy_2019, Jenks:2023pmk}. Under the current framework, this corresponds to the case of vanishing expansion coefficients except a non-zero $\rho$ in our parameterization. As a result, there is only velocity birefringence, but not amplitude birefringence, meaning $\delta\phi_A=0$ in Eq.~(\ref{eq:PV_parameterization}). The velocity birefringence changes the relative phase between the chiral states, and hence causes modifications in the observed polarization content after propagation relative to the GR prediction. However, the frequency-independent $\delta\phi_V$ is degenerate with both the source redshift and the polarization angle of the source~\cite{Jenks:2023pmk}. The degeneracy with redshift is clear from Eq.~(\ref{eq:PV_parameterization}), while the degeneracy with the polarization angle is realized as 
\begin{align}
    \psi\rightarrow \psi+\frac{\delta\phi_V}{2},
\end{align}
which reassigns the shifts in the polarization mixture from propagation to emission. Measuring the polarization angle precisely is challenging at present~\cite{Charisi:2021dwc}, which makes constraining such a theory under the presented framework difficult in practice.

\section{Injection and recovery}
\label{sec:injection_and_recovery}
We now demonstrate the cross-correlated CW framework numerically by injecting beyond-GR signals into simulated PTA data and recovering them with Bayesian inference, following the approach of~\citet{mingarelli2026fingerprintsindividualsupermassiveblack}. We consider two representative cases drawn from Sections~\ref{sec:non-tensor} and~\ref{sec:dispersion}: a quadrupolar breathing mode and a massive graviton. For each, we compare the cross-correlated beyond-GR model to three alternatives: a pure-GR CW model, an uncorrelated CW model, and a Hellings-Downs-correlated GWB model.

\begin{figure*}
    \centering
    \includegraphics[width=0.5\linewidth]{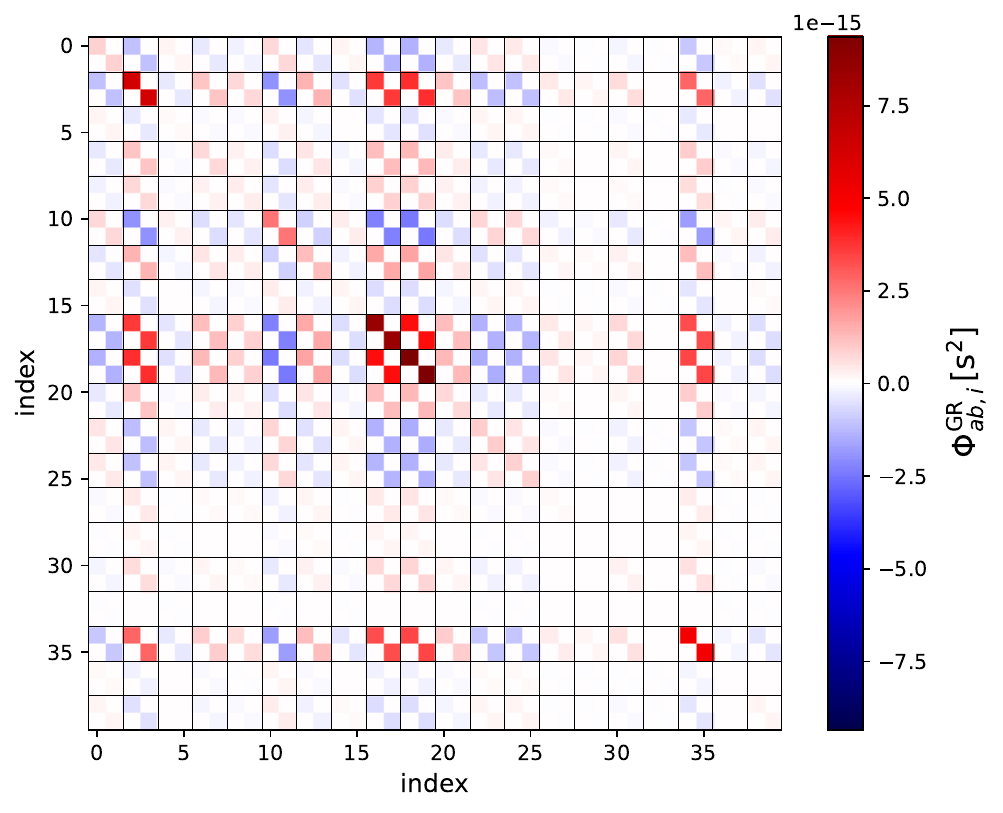}(a)\\
    \includegraphics[width=0.47\linewidth]{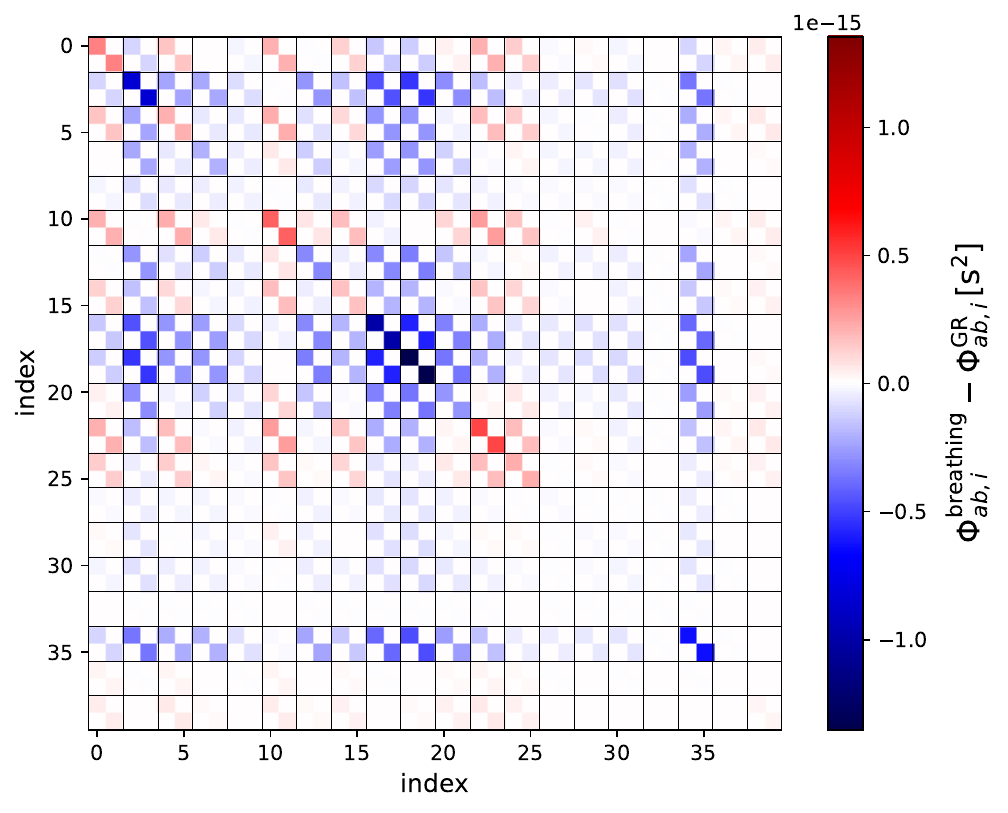}(b)
    \includegraphics[width=0.47\linewidth]{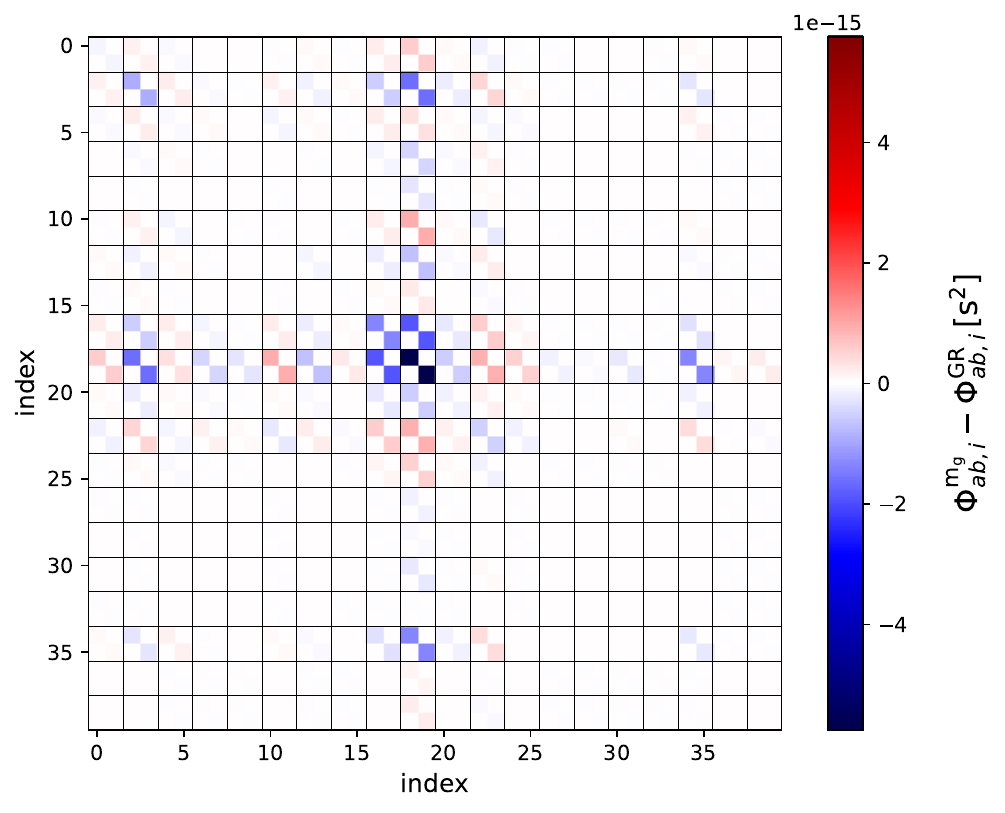}(c)
    \caption{The beyond-GR cross-correlation matrix (the off-diagonal terms) offers the numerical basis for discerning the modifications to GR, explicitly defined in Eq.~(\ref{eq:Phi_matrix_CW}). On top, (a) is the pure GR $\Phi^{GR}_{ab, i}$ for the first 20 pulsars in the simulated data. At the bottom are the corrections due to the beyond-GR effects. (b) is the breathing mode induced correction to the inter-pulsar correlation with a CW at $f_{\rm gw}=6\, \rm nHz$, while (c) shows the massive graviton induced corrections for a CW with the same parameters except for the included beyond-GR effects. The breathing mode modification produces an average of $9.1\%$ deviation in the matrix entries from the GR case, while massive gravity $37.8\%$. These two non-GR components have rather different patterns on the inter-pulsar correlation modifications, making the numerical differentiation possible.}
\label{fig:Phi_matrix_diff}
\end{figure*}

We use \texttt{pta\_replicator}~\cite{Becsy:2025:pta_replicator} to simulate an idealized array of $N_{\rm psr}=100$ pulsars, placed randomly on the sky following a uniform sky distribution, with equal observation timespans, uniform cadence, and identical time of arrival (TOA) uncertainties of $100\, \rm ns$. A CW signal containing beyond-GR physics is then injected, with parameters listed in Table~\ref{table: parameters}. We set the strain amplitude of the source as $5 \times 10^{-15}$, slightly below the current upper limit on CW strain amplitude~\cite{Agazie_2023}. 

As in \citet{mingarelli2026fingerprintsindividualsupermassiveblack}, for our recovery analysis we search for no additional noise or signals besides the CW and marginalize over a simple 5-parameter pulsar timing model for each pulsar. We adopt the standard Gaussian PTA likelihood, with data covariance
\begin{equation}
    \bm{C} = \bm{D} + \bm{F}\bm{\Phi}\bm{F}^{T},
\end{equation}
where $\bm{D}$ packages the timing model and the TOA uncertainty contributions \citep{Johnson+2024}. $\bm{F}$ is the CW spectral design matrix \citep{mingarelli2026fingerprintsindividualsupermassiveblack}, constructed as a sine–cosine pair for every pulsar. Labeling pulsars by $k=1,\dots,N_{\rm psr}$, the two CW basis functions for pulsar $k$ occupy columns $2k-1$ and $2k$:
\begin{align}
    F_{i,2k-1} &= \cos\bigl(2\pi f_{\rm gw} t_{i,{\rm PSR}_k}\bigr),\\
    F_{i,2k}   &= \sin\bigl(2\pi f_{\rm gw} t_{i,{\rm PSR}_k}\bigr),
\end{align}
with $i$ running over all TOAs in pulsar $k$. 

The CW spectral covariance matrix is modeled as
\begin{equation}
    \Phi_{ab,j} = \Gamma^{\rm theory}_{ab}(\hat{\bm{\Omega}}_{\rm gw})
    \frac{h_0^2}{4\pi^2 f_{\rm gw}^2},
    \label{eq:Phi_matrix_CW}
\end{equation}
where the overall amplitude is set by $h_0^2/(4\pi^2 f_{\rm gw}^2)$, and $\Gamma^{\rm theory}_{ab}(\hat{\bm{\Omega}}_{\rm gw})$ depends on the specific modified theory of gravity.

We consider an explicit example corresponding to each of Sec.~\ref{sec:non-tensor} and Sec.~\ref{sec:dispersion} independently. For the non-tensorial polarizations, we consider a quadrupolar breathing mode on top of the transverse tensor modes. For the dispersion relations, we consider a generic theory with massive graviton. As shown in Sec.~\ref{sec:birefringence}, most birefringence effects are impractical to detect in the PTA band at present, hence we do not include an injection and recovery here. Throughout this section we use \texttt{enterprise} to construct the likelihood functions and prior distributions, and the \texttt{nautilus} importance nested sampling code to generate posterior samples~\citep{nautilus}.

There are some key questions we seek to probe in this injection-and-recovery study: i) how well does the beyond-GR cross-correlated CW model describe the data versus an uncorrelated model, and distinguish the beyond-GR CW signal from a GWB signal? ii) does the cross-correlated CW model recover the beyond-GR parameters with reasonable statistical significance and accuracy? iii) is it justified to assume GR when first making a detection of the CWs, or is it necessary to include the beyond-GR effects in the model to avoid bias?

\subsection{Breathing mode}
As an example of measuring non-tensorial polarization states, we inject a breathing mode in addition to the transverse tensor polarization states of GR. We follow the parametrized framework for scalar-tensor inspiral GWs from binary coalescences introduced in~\cite{Takeda_2022}. As a proof of concept, we restrict to a generic quadrupolar breathing mode that propagates at $c$, but a more comprehensive search could include the dipolar breathing mode~\cite{Will:1977bb, Yunes_2016, Chatziioannou:2012rf}, other non-tensorial polarization modes, as well as a modified propagation speed of the alternative polarizations~\cite{Schumacher:2025plq, Schumacher:2023jxq}. The polarization content is thus
\begin{align}
\label{eq: breathing_mode_waveform}
h_+(t)&=h_0(1+\cos^2\iota)\cos(\omega_0 t+\phi_0)\notag\\
h_{\times}(t)&=-2h_0\cos\iota\sin(\omega_0t+\phi_0)\notag\\
h_{B}(t)&=-2h_0\mathcal{A}\sin^2\iota\cos(\omega_0 t + \phi_0).
\end{align}
The generic parameter $\mathcal{A}$ describes the relative amplitude of the breathing mode as compared to the polarization states of GR. Following Eq.~(\ref{eq:master_correlation}), the cross correlation in the cosmic rest frame is
\begin{widetext}
    
\begin{align}
\label{eq:breathing_mode_cross_correlation}
    &\Gamma^{\mathrm{Breathing}}_{ab}=
    \frac{1}{2}\left[1 + 6\cos^2\iota + \cos^4\iota\right](F^+_aF^+_b + F^\times_aF^\times_b) 
+ \frac{1}{2}\sin^4\iota\cos(4\psi)(F^+_aF^+_b - F^\times_aF^\times_b) \nonumber\\ 
    &+\frac{1}{2}\sin^4\iota\sin(4\psi)(F^+_aF^\times_b + F^\times_aF^+_b) -2\mathcal{A}(1-\cos^4\iota)\Big[\cos2\psi F_a^+ F_b^B + \sin2\psi F_a^{\times} F_b^B\Big] \nonumber\\
    &-2\mathcal{A}(1-\cos^4\iota)\Big[\cos2\psi F_a^B F_b^+ + \sin2\psi F_a^B F_b^{\times}\Big]
    +4\mathcal{A}^2 (1-\cos^2\iota)^2 F_a^B F_b^B,
\end{align}
\end{widetext}
where all the antenna patterns are defined in the cosmic rest frame accordingly. For comparison, we also consider an uncorrelated diagonal CW model as a conservative null hypothesis, as the mismodeled intrinsic pulsar noise only affects the autocorrelations. This is analogous to a \emph{common uncorrelated red noise} (CURN) model in GWB searches. In this model, only the autocorrelations are included, and the polarization and inclination angles are marginalized over to give the conventional numerical factors under our normalization
\begin{equation}
    \Gamma^{\rm breathing}_{ab, \rm UC}= \frac{8}{5}\delta_{ab}\Bigg[(F_a^+)^2+(F_a^{\times})^2+\frac{4}{3}\mathcal{A}^2(F_a^B)^2\Bigg],
    \label{eq:Gamma_breathingauto}
\end{equation}
To understand the impact of model mis-specification, we also introduce a model where the CW is recovered as a Hellings-Downs correlated process, representative of an isotropic GWB, instead of the correct CW model. With the breathing mode included, the cross-correlation is:
\begin{equation}
    \Gamma^{\rm Breathing}_{ab, \rm GWB}= \Gamma_{ab}^{\rm HD}+\mathcal{A}^2\Gamma_{ab}^{\rm B},
    \label{eq:Gamma_breathingGWB}
\end{equation}
where 
\begin{subequations}
\begin{align}
    \Gamma^{\rm HD}_{ab}
&= \frac{3}{2}x_{ab}\ln x_{ab} - \frac{x_{ab}}{4} + \frac{1}{2}(1+\delta_{ab}),\\
\Gamma^{\rm B}_{ab}
&= \frac{1}{8}(3+\beta_{ab})+\frac{\delta_{ab}}{2},\\
x_{ab}&=\frac{1-\beta_{ab}}{2},\\
\beta_{ab}&=\cos\zeta_{ab}.
\end{align}
\end{subequations}
$\zeta_{ab}$ here is the angular separation between pulsar a and b. We have assumed that the physical relative amplitude of the breathing mode is the same for a CW and the GWB to make a controlled comparison. 
\begin{table}
    \centering
    \begin{tabular}{ c | c @{}}
        \hline\hline Dataset property & Description \\
        \hline $N_{\rm psr}$ & $100$ \\
        \hline $T_{\rm obs}$ & $16.42$ yr \\
        \hline $N_{\rm TOA}$ & $300$ \\
        \hline $\sigma_{\rm TOA}$ & $100$ ns \\
        \hline $L_p$ & $1$ kpc \\
        \hline\hline Injected CW parameter & Value \\
        \hline $\theta_{\rm gw}$ & $5\pi/7$ \\
        \hline $\phi_{\rm gw}$ & $5\pi/3$ \\
        \hline $\mathcal{M}_c$ & $1.58\times10^9$ M$_\odot$ \\
        \hline $h_0$ & $5\times 10^{-15}$ \\
        \hline $f_{\rm gw}$ & $6$ nHz\\
        \hline $\Phi_0$ & $0$ \\
        \hline $\psi$ & $\pi/6$ \\
        \hline $\iota$ & $\pi/2$ \\
        \hline $\mathcal{A}$ & $3\times10^{-2}$ \\
        \hline $m_g$ & $1\times10^{-23}\, \rm eV$ \\
        \hline\hline
    \end{tabular}
    \caption{Simulated dataset and injected CW properties. The injected CW includes pulsar terms with random pulsar phases and the pulsar distances fixed to 1 kpc. The beyond GR signals are injected and analyzed independently for the breathing mode and massive gravity.}
    \label{table: parameters}
\end{table}
Breathing mode has been searched for across the GW spectrum~\cite{Wu_2022, NANOGrav:2023ygs, LIGOScientific:2021sio, Takeda_2022}, with both GWB and GWs from individual sources. There is no universal limit on the amplitude of the breathing mode, as the relative amplitude of the breathing mode versus the tensorial modes depends on the properties of the particular binary source, such as the scalar charge or scalar hair. In~\cite{Takeda_2022}, such limits are extracted for GW 170814 and GW 170817, which are the most stringent constraints in the strong-field regime near merger. We quote the better limit from GW 170817, $R_{\rm ST}\lesssim 0.034$, where $R_{\rm ST}$ is defined in Eq. (11) of \cite{Takeda_2022}. For our choice of $\iota=\pi/2$, this translates to $\mathcal{A}\lesssim0.034$. We then take $\mathcal{A}=3\times10^{-2}$ as a benchmark value for the injected signal.

\begin{figure*}
    \centering
    \includegraphics[width=0.49\linewidth]{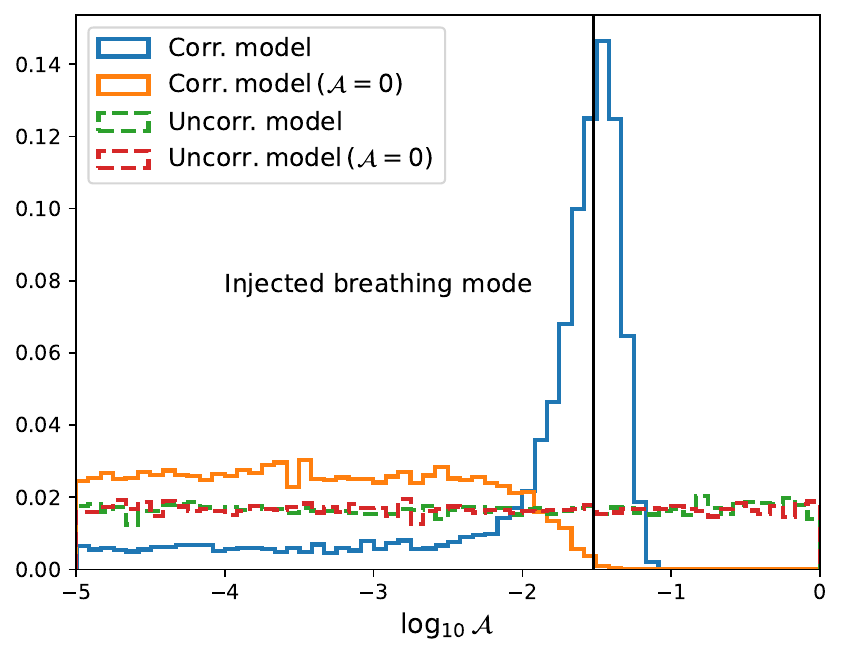}
    \includegraphics[width=0.483\linewidth]{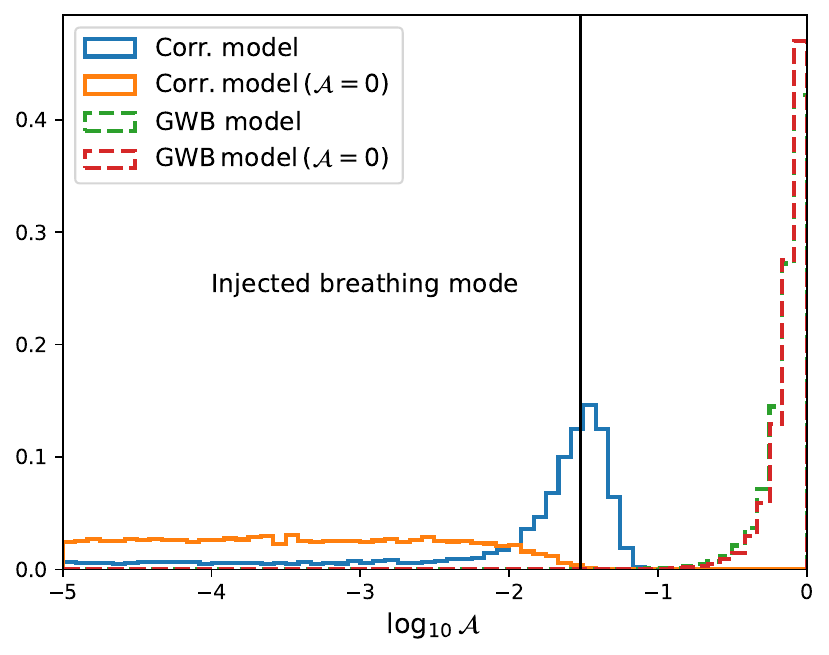}
    \caption{The cross-correlated CW model is favored by the data over both the uncorrelated CW model (top) and the GWB model (bottom), with GR + breathing mode. The Bayes factors are $\log \mathcal{B}_{UC, B}^{CC}=68.9$ and $\log \mathcal{B}_{\rm GWB, B}^{CC}=117.2$ respectively. \textit{Top}: A comparison between the $\mathcal{A}$ posteriors from the cross-correlated model and the uncorrelated diagonal model defined in Eq.~(\ref{eq:Gamma_breathingauto}). The cross-correlated model recovers the injected $\mathcal{A}$ and sets constraints on $\mathcal{A}$ when the injected breathing mode is zero, while the uncorrelated model results in a uniform posterior. \textit{Bottom}: A comparison between the cross correlated CW model and the GWB model defined in Eq.~(\ref{eq:Gamma_breathingGWB}). The GWB model produces a posterior densely peaked near $\mathcal{A}\sim 1$ for both data, suggesting a model misspecification.}
\label{fig:breathing_mode_posterior}
\end{figure*}

In the left panel of Fig.~\ref{fig:Phi_matrix_diff}, we show the change in the correlation matrix defined in Eq.~(\ref{eq:Phi_matrix_CW}) by the breathing mode. This change is the numerical basis for discerning the deviation from GR. Fig.~\ref{fig:breathing_mode_posterior} shows the posterior distributions of the relative amplitude of the breathing mode $\mathcal{A}$ as defined in Eq.~(\ref{eq: breathing_mode_waveform}).

We generate two datasets: one with a breathing mode injected and one with only the GR signals, which we call the null data. The data realized with the parameters in Table~\ref{table: parameters} have $\rm SNR\approx64$, calculated using the $\textit{\textbf{D}}$ matrix~\cite{mingarelli2026fingerprintsindividualsupermassiveblack, O_Beirne_2019}. We adopt uniform priors on seven parameters defined below,
\begin{subequations}
\label{eq:priors}
\begin{align}
    \log_{10} \mathcal{A} &\sim \mathcal{U}(-5,  0), \label{eq:priors_1}\\
    \log_{10} h_0 &\sim \mathcal{U}(-18,-10), \\
    \log_{10} f_{\rm gw} &\sim \mathcal{U}(-9,-8), \\
    \cos\theta_{\rm gw} &\sim \mathcal{U}(-1,1), \\
    \phi_{\rm gw} &\sim \mathcal{U}(0,2\pi), \\
    \cos\iota &\sim \mathcal{U}(-1,1), \\
    \psi &\sim \mathcal{U}(0,\pi).
\end{align}
\end{subequations}

The parameter $\psi$ does not sweep the full cycle of $(0, 2\pi)$ in the prior due to the degeneracy in the cross correlation (Eq.~(\ref{eq:breathing_mode_cross_correlation})). We carry out a systematic comparison among the three models derived above, with both breathing mode injected data and null data: the cross correlated CW model (Eq.~(\ref{eq:breathing_mode_cross_correlation})), the uncorrelated CW model (Eq.~(\ref{eq:Gamma_breathingauto})), and the GWB model (Eq.~(\ref{eq:Gamma_breathingGWB})).
The posteriors in the top panel of Fig.~\ref{fig:breathing_mode_posterior} demonstrate that the cross-correlated model derived in this work correctly recovers the breathing mode (blue line), and places an upper limit on $\mathcal{A}$ when using the null data (orange line). In contrast, the posteriors from the uncorrelated CW model do not deviate from the uniform prior, for both data with signal injected (blue solid line) and the null data (orange solid line), suggesting an inability to extract information about the breathing mode from the pulsar autocorrelations. The Bayes factor of the cross-correlated model over the uncorrelated model is $\log \mathcal{B}_{UC, B}^{CC}=68.9$, strongly favoring the cross-correlated model (see Appendix~\ref{app:Bayes_factors}). This comparison indicates the advantage that CW cross correlation brings to the search for beyond-GR polarizations. In the bottom panel of Fig.~\ref{fig:breathing_mode_posterior}, we compare the cross correlated CW model with the GWB model. For both data, the GWB model results in posteriors of $\mathcal{A}$ peaked near $10^0$, two orders of magnitude larger than the injected breathing mode. This shows that attempting to treat a CW as a Hellings-Downs correlated process may lead to spurious evidence for a breathing mode. Nonetheless, a Bayes factor of $\log \mathcal{B}_{\rm GWB, B}^{CC}=117.2$ indicates the CW signal is unlikely to be confused with a GWB. Table~\ref{tab:bgr_bayes_factors} compiles the log Bayes factors for the model comparisons.
\begin{table*}[t]
\centering
\begin{tabular}{|l||c|c|c|c|}
\hline\hline
\hspace{2.8cm} & \multicolumn{4}{c|}{\textbf{$\log\mathcal{B}$ for the beyond-GR cross-correlated model vs. reference models}} \\
\hline
\textbf{Beyond-GR hypothesis} 
& \makebox[2.8cm]{\textbf{Uncorr.\ CW}}
& \makebox[2.8cm]{\textbf{Corr.\ GWB}}
& \makebox[2.8cm]{\textbf{Corr.\ GR CW}}
& \makebox[2.8cm]{\textbf{Corr.\ GR CW}}\\
\hline
\textbf{$\sigma_{\rm TOA}$} 
& 100 ns & 100 ns & 100 ns & 10 ns \\
\hline\hline
\textbf{Breathing mode} & 68.9 & 117.2 & 1.1 & 3.0 \\
\hline
\textbf{Massive gravity} & 68.3 & 251.3 & 5.3 & 10.5 \\
\hline\hline
\end{tabular}
\caption{
\textit{Log Bayes factors} for the two beyond-GR hypotheses under four model comparisons. The \textit{Uncorr. CW} and \textit{Corr. GWB} models are strongly disfavored given the injected signals. The Log Bayes factors over the \textit{Corr. GR CW} model 
encodes how strongly the beyond-GR hypothesis is preferred. At $\sigma_{\rm TOA}=100\, \rm ns$, the breathing mode is moderately favored while massive gravity is strongly supported by the log Bayes factor, given the injected beyond-GR parameters. With $\sigma_{\rm TOA}=10\, \rm ns$, both Bayes factors increase, with the \textit{log Bayes factor} for the breathing mode also reaching 3.0, suggesting evidence for the breathing mode. }
\label{tab:bgr_bayes_factors}
\end{table*}

Strategically, while we can include the beyond-GR effects in the model as we search for the CW sources, it is both computationally expensive and unlikely that we could account for all the potential deviations from GR exhaustively, due to the limits of our knowledge and possible model misspecification. Consequently, it is instructive to understand whether a cross correlated CW model that assumes pure GR is able to identify the CW sources with reasonably accurate recovery of their parameters, even when real deviations from GR are present in the data. Here we apply the pure GR cross correlated CW model introduced in~\cite{mingarelli2026fingerprintsindividualsupermassiveblack} to our simulated data containing the breathing mode. As shown in Fig.~\ref{fig:full_posterior_breathing_mode} (Appendix~\ref{app: full_posteriors}), we find that the GR model is still able to recover the sky location, frequency, strain amplitude, polarization and inclination angle, with the injected breathing mode present in the data. Including the breathing mode introduces only a slight sharpening of the constraint on inclination angle. Although the Bayes factor $\log\mathcal{B}_{\rm B}=1.1$ represents only weak preference for the breathing-mode model over GR, the GR cross-correlated model still recovers the source parameters---sky location, frequency, strain amplitude, polarization, and inclination---without significant bias, supporting the use of a GR template for initial CW detection even when a breathing mode is present. For the null data, the posterior on $\mathcal{A}$ yields an upper limit with no peak; all other parameters are accurately recovered. The full joint posteriors are shown in Appendix~\ref{app: full_posteriors}.

We emphasize that the Savage-Dickey Bayes factors reported here should not be interpreted as evidence for or against a detection. The value $\log\mathcal{B}_B\approx 1.1$ corresponds to the alternative hypothesis being only about three times more likely than the null, which does not constitute strong evidence. Higher SNR improves detection confidence: repeating the analysis with $\sigma_{\rm TOA}=10$~ns (with all other parameters held fixed) recovers the breathing mode with $\log\mathcal{B}_B=3.0$, and the posterior peak at the injected $\mathcal{A}$ is visibly sharper (Appendix~\ref{app: recovery_SNRs}). More fundamentally, in the context of these injection-recovery tests, the Savage-Dickey Bayes factors serve only as a diagnostic of whether the pipeline prefers the signal model over the null model for a given data realization. A statistically robust interpretation as a detection metric would require systematic calibration on a large ensemble of simulated datasets, characterizing the Bayes factor distributions under both signal and noise hypotheses to establish detection thresholds and false-alarm probabilities. Such a calibration lies beyond the scope of this work and will be addressed in future studies incorporating realistic noise environments.

\subsection{Massive gravity}
Massive gravity has been studied extensively across modified-gravity theories, following the original Fierz-Pauli action~\cite{FierzPauli1939, Will_1998, Rubakov_2008}. Limits on the graviton mass have been derived from GW interferometry~\cite{LIGOScientific:2021sio, baka2025testinggeneralrelativitygravitational}, PTAs~\cite{Wang_2024, Wu_2024}, large galaxy surveys~\cite{Desai_2018, loeb2024newlimitgravitonmass}, and solar-system dynamics~\cite{PhysRevLett.123.161103}. Indirect constraints reach values as low as $5\times10^{-32}$~eV, set by demanding that a graviton-mass-induced Yukawa correction to the gravitational potential remain within $1\sigma$ of the convergence measured between the CMB dipole anisotropy and the 2MASS matter distribution~\cite{loeb2024newlimitgravitonmass}. This is roughly eight orders of magnitude tighter than current direct GW constraints. The tightest direct upper limit from PTAs is $8.2\times10^{-24}$~eV~\cite{Wu_2024}, with ground-based detectors giving a slightly weaker $2.42\times10^{-23}$~eV~\cite{LIGOScientific:2021sio}. We inject a graviton mass of $1\times10^{-23}$~eV, just below current direct GW upper limits, to test recovery at the edge of present sensitivity.

We use the Earth term cross correlation derived in Eq.~(\ref{eq:dispersion_correlation}) as the detection scheme, where the antenna patterns now read
\begin{align}
\label{eq:massive_graviton_antenna_patterns}
    F_a^{P}(\hat{\bm{\Omega}})
&= \frac{1}{2}
\frac{\hat p_a^i \hat p_a^j}{1+\chi\,\hat{\Omega}\!\cdot\!\hat{\bm{p}}_a}\,
e^{P}_{ij}(\hat{\bm{\Omega}}),\notag\\
\chi&=\sqrt{1-\frac{m_g^2c^4}{\omega^2\hbar^2}},
\end{align}
where we have restored the $\hbar$ from discussion in Section~\ref{sec:dispersion}. GR has only 2 degrees of freedom, whereas massive gravity has 5 degrees of freedom. We assume that the non-tensorial polarization states are suppressed, and focus on the effects of the modified dispersion relation for clarity. There is no practical barrier that prevents a general search with both modified dispersion and alternative polarization states. The cross correlation in the cosmic rest frame becomes
\begin{align}
\label{eq:massive_graviton_cross_correlation}    &\Gamma^{\mathrm{m_g}}_{ab}=\notag\\
    &\frac{1}{2}\left[1 + 6\cos^2\iota + \cos^4\iota\right](F^+_aF^+_b + F^\times_aF^\times_b) \nonumber \\
    &+ \frac{1}{2}\sin^4\iota\cos(4\psi)(F^+_aF^+_b - F^\times_aF^\times_b) \nonumber \\
    &+\frac{1}{2}\sin^4\iota\sin(4\psi)(F^+_aF^\times_b + F^\times_aF^+_b),
\end{align}
where all the antenna patterns are defined according to Eq.~(\ref{eq:massive_graviton_antenna_patterns}). Similar to the breathing mode, an uncorrelated CW model serving as the null hypothesis can be constructed for massive gravity
\begin{equation}
    \Gamma^{\rm m_g}_{ab, \rm UC}=\frac{8}{5} \delta_{ab}\Bigg[(F_a^+)^2+(F_a^{\times})^2\Bigg].
    \label{eq:Gamma_massive_gravityauto}
\end{equation}
The massive-gravity GWB cross correlation has been derived in~\cite{Lee_2010, Liang:2021bct, Wu_2023}. A complete evaluation of this ORF requires numerical integration for each specific realization of the GW frequency, graviton mass, and pulsar pair. It is therefore computationally expensive in a Bayesian analysis and, moreover, requires explicit knowledge of the pulsar distances; see, for example, Eq.~(29) of Ref.~\cite{Liang:2021bct}. To avoid this computational bottleneck, we use the fact that in the limit $f_{\rm gw}d_a\gg1$, the ORF can be accurately approximated by dropping the rapidly oscillating pulsar-term exponential factors~\cite{Liang:2021bct}. For our simulated data, $\chi\simeq0.92$ and $f_{\rm gw}d_a\simeq620$, so this approximation is well justified. The massive-gravity GWB cross correlation is then approximated by~\cite{Wu_2023, Liang:2021bct}
\begin{equation} 
\label{eq: massive_gravity_GWB_integral} \Gamma^{\mathrm{\rm m_g}}_{ab, \rm GWB}=\frac{1}{8\pi}\sum_{P=+,\times}\int_{S^2}d\hat{\Omega}\, F_a^P(\hat{\Omega})F_b^P(\hat{\Omega}), 
\end{equation}
where the antenna pattern functions are defined in Eq.~(\ref{eq:massive_graviton_antenna_patterns}).

We generate the signal-injected data and null data, with a $\rm SNR\approx63$. The modification of the correlation $\Phi_{ab,j}$ matrix due to the massive graviton is quite different from that due to the breathing mode, as shown in the right panel of Fig.~\ref{fig:Phi_matrix_diff}. Here we have kept everything including the CW frequency the same except the beyond-GR injections. Similar to the breathing mode case, the data are analyzed with the cross correlated CW model, the uncorrelated CW model and the GWB model respectively. The priors are mostly the same as in the breathing mode case, except that the prior on $\mathcal{A}$ is replaced by a prior on $m_g$, and the prior range of $\psi$ is further shrunk due to the four-fold degeneracy in Eq.~(\ref{eq:massive_graviton_cross_correlation}): 
\begin{subequations}
\label{eq:priors_mg}
\begin{align}
    \log_{10} m_g &\sim \mathcal{U}(-11,  -8.2), \\
    \psi &\sim \mathcal{U}(0,\pi/2).
\end{align}
\end{subequations}
The prior on $m_g$ is set in Hz for convenience and the upper cut-off comes from the assumption that the GWs propagate and do not attenuate in the PTA range.

\begin{figure}[ht!]
    \centering
    \includegraphics[width=1\linewidth]{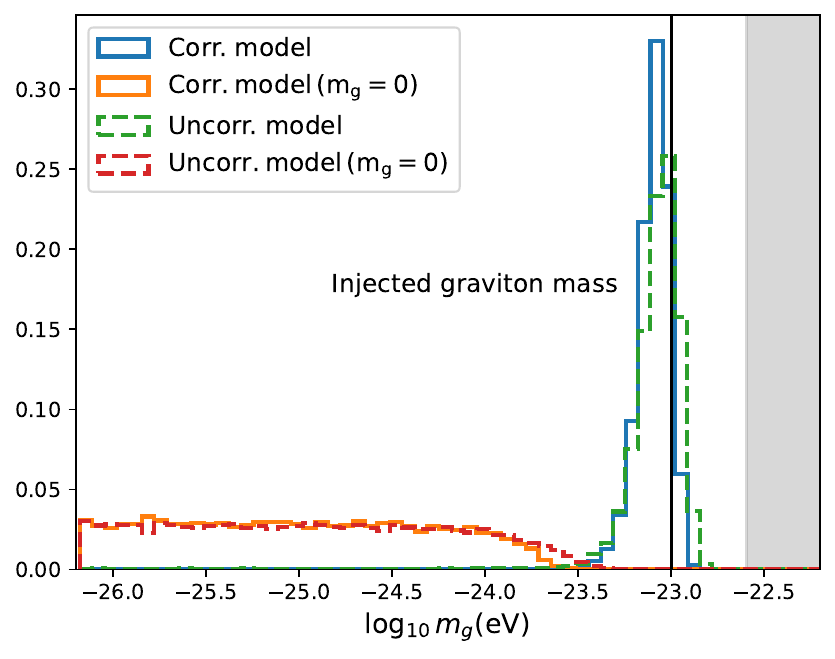} \includegraphics[width=1\linewidth]{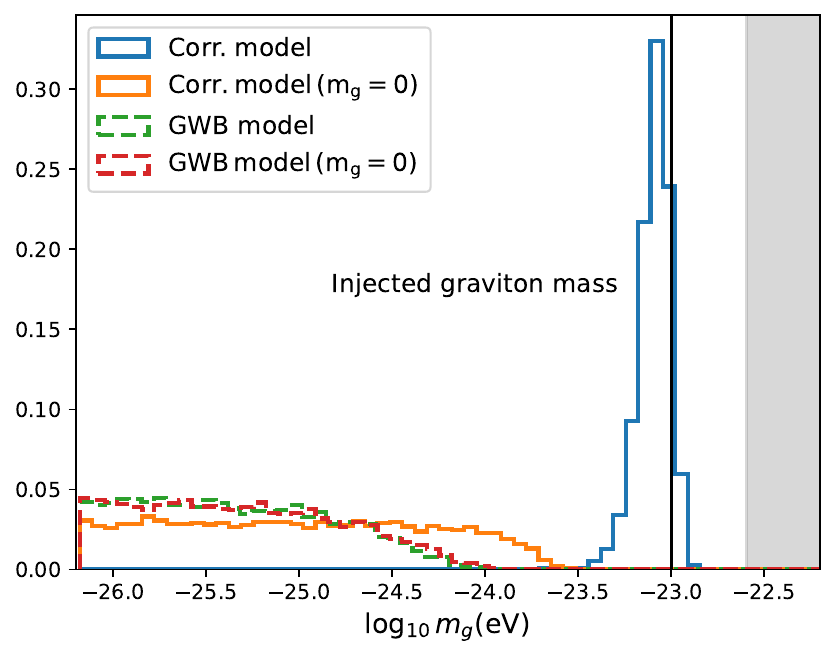}
    \caption{The cross-correlated CW model is preferred by the data over the uncorrelated CW model (top) and the GWB model (bottom) for the data with massive gravity, with a Bayes factor of $\log \mathcal{B}_{UC, m_g}^{CC}=68.3$ and $\log \mathcal{B}_{\rm GWB, m_g}^{CC}=251.3$ respectively. \textit{Top}: a comparison between the $m_g$ posteriors from the cross-correlated model and uncorrelated model defined in Eq.~(\ref{eq:Gamma_massive_gravityauto}). For zero graviton mass, the uncorrelated CW model is able to constrain the graviton mass almost as well as the cross-correlated model. The two models also result in similar posterior distributions when injected $m_g=0$. Nevertheless, the Bayes factor strongly prefers the cross-correlated model. The shaded region corresponds to GW attenuation, where $\chi$ becomes imaginary (Eq.~\ref{eq:massive_graviton_antenna_patterns}). \textit{Bottom}: Comparison between the cross-correlated CW model and the massive-gravity GWB cross-correlation model. The GWB model does not recover the injected graviton mass, instead favoring smaller graviton masses regardless of whether a massive graviton is present in the data. This indicates that the GWB model mischaracterizes the CW signal in the data.}
    \label{fig: mg_CC_UC}
\end{figure}

First, we compare the cross-correlated (Eq.~(\ref{eq:massive_graviton_cross_correlation})) and uncorrelated CW models (Eq.~(\ref{eq:Gamma_massive_gravityauto})), the posteriors of which are presented in the top panel of Fig.~\ref{fig: mg_CC_UC}. Both models are able to set limits on the graviton mass with the null data, and generate a posterior of $m_g$ around the injected value. This is possibly due to the fact that the graviton mass induced change in the antenna pattern functions is common to both the cross-correlated and uncorrelated CW models, and cannot be attributed to other parameter shifts due to degeneracies. Nevertheless, the Bayes factor favoring the cross-correlated model over the uncorrelated model is $\log \mathcal{B}_{UC, m_g}^{CC}=68.3$, suggesting again a strong preference for the cross-correlation. Second, we compare the cross-correlated CW model with the massive-gravity GWB model, whose posteriors are shown in the bottom panel of Fig.~\ref{fig: mg_CC_UC}. The GWB model produces similar posteriors for $m_g$ in both the massive-gravity injection and the massless case. Mismodeling the CW signal as a GWB thus sets a spurious upper limit on the graviton mass, much lower than the actual injection. The Bayes factor in favor of the cross-correlated CW model over the GWB model is $\log \mathcal{B}_{\rm GWB, m_g}^{\rm CC}=251.3$, further indicating that the GWB model is misspecified for a CW signal in massive gravity. Therefore, a potential detection of a massive graviton through a CW signal is unlikely to be mistaken for a GWB signal. Third, we apply the GR cross correlated CW model in~\cite{mingarelli2026fingerprintsindividualsupermassiveblack} on the data containing a massive graviton. The GR model again recovers the source parameters fairly well, suggesting the feasibility of a GR-only model in the first detection of CW sources. Including the effect of a massive graviton yields only slight improvements to the recovery of $h_0$ and $\psi$, as shown in Fig.~\ref{fig:full_posterior_massive_graviton} (Appendix~\ref{app: full_posteriors}). The massive gravity cross-correlated model is still favored with a Bayes factor of $\log\mathcal{B}_{m_g}=5.3$. Finally, just as in the case of breathing mode, we find that a higher SNR realized by a smaller $\sigma_{\rm TOA}=10\, \rm ns$ improves the recovery of $m_g$ with a higher Bayes factor of $\log\mathcal{B}_{m_g}=10.5$, as shown in Appendix~\ref{app: recovery_SNRs}.

Although not included in the current pipeline, it is also worthwhile to discuss the detectability of the phase delay with pulsar terms for future searches. By Eq.~(\ref{eq:dispersion_phase}) and Eq.~(\ref{eq: dispersion_relation}), the phase delay induced by the graviton mass is
\begin{align}
\Delta\Phi^{\rm Disp}_a
&= \frac{m_g^2c^3d_a}{2\omega\hbar^2}\, \hat\Omega\!\cdot\!\hat p_a\,.
\end{align}
An order of magnitude estimate with nHz GWs, kpc pulsar distances at the injected graviton mass results in a phase shift of $\mathcal{O}(10^{-1})$. It also clarifies that the phase delay scales quadratically with the graviton mass, in addition to the geometric dependence determined by the $\hat\Omega\!\cdot\!\hat p_a$ term.

\begin{figure}
    \centering
    \includegraphics[width=1\linewidth]{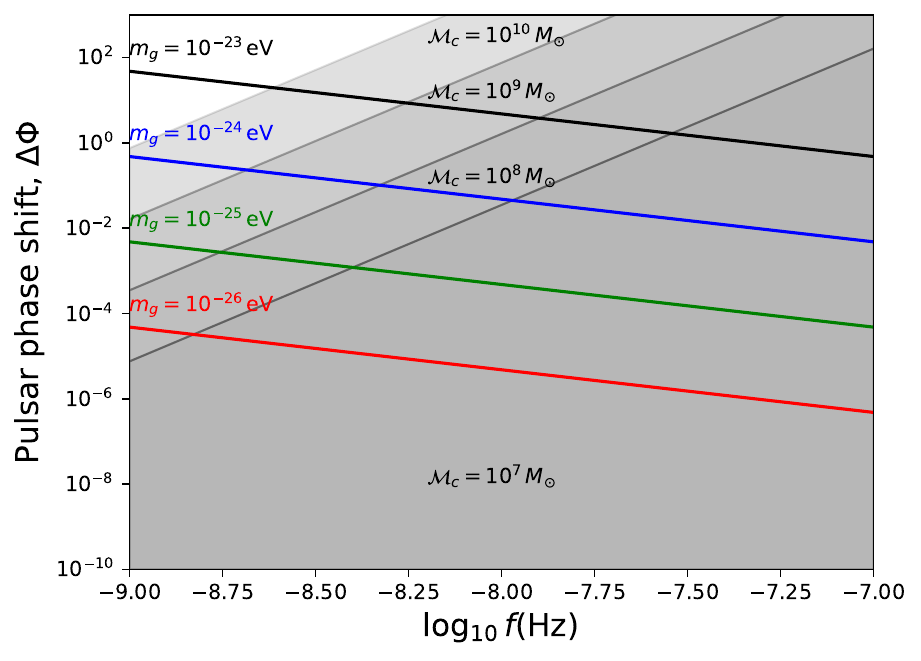}
    \caption{On top of the cross correlation, the pulsar term phase shift offers a unique probe of modified dispersion relations such as massive gravity. Here we plot massive gravity induced phase delays as a function of GW frequency, for different graviton masses. The shaded regions are the estimated phase correction due to the GW frequency evolution of the source. Here we assume a typical MSP pulsar distance of 1 kpc, which can be understood as the arm length of the detector in a more general sense. The lower the frequency is, the larger the graviton mass induced phase shift and the smaller the frequency evolution confusion signal is. }
\label{fig:graviton_mass}
\end{figure}

It is worth estimating how much the SMBHB evolution would contribute to the phase delay and confuse with a modified dispersion such as that due to massive gravity considered here. The GW frequency evolves as
\begin{equation}
    \dot{f}_{gw}=\frac{96}{5}\pi^{8/3}\mathcal{M}_c^{5/3}f_{\rm gw}^{11/3}.
\end{equation}
By the slow evolution of the SMBHB at low frequencies, the phase correction due to the frequency evolution can be estimated as
\begin{equation}
    \Delta\Phi_{\rm evol}=2\pi\int_0^{T_c}\dot{f}_{gw}t\, dt=\pi \dot{f}_{gw} T_c^2
\end{equation}
where $T_c$ is the light travel time from the pulsar to Earth. For our choices of parameters listed in Table~\ref{table: parameters} and taking $f_{\rm gw}$ as the frequency of the GW at Earth, $\Delta\Phi_{\rm evol}\sim \mathcal{O}(10^1)\, \rm rad$, which completely overshadows the phase delay induced by the graviton mass. If the GWs pass the pulsar before arriving at Earth, $\Delta\Phi_{\rm evol}$ is underestimated; otherwise $\Delta\Phi_{\rm evol}$ flips the sign and its magnitude is overestimated. For lower chirp mass and frequency, $\Delta\Phi_{\rm evol}$ could be sub-dominant to $\Delta\Phi^{\rm Disp}_a$. For example, for $\mathcal{M}_c=10^8 M_\odot$ and $f_{\rm gw}=1\, \rm nHz$, $\Delta\Phi_{\rm evol}\sim \mathcal{O}(10^{-4})\, \rm rad$. Fig.~\ref{fig:graviton_mass} shows a comparison of the graviton mass induced phase delays and frequency evolution induced phase drifts, for various graviton masses and SMBHB masses. The lower the frequency is, the more sensitive the signal is to any potential massive graviton induced modification to the dispersion relation of GWs.

\section{Conclusions and Discussions}
\label{sec:discussions}
We have constructed a unified framework for testing theories of gravity using CWs from individual SMBHBs in the low-frequency regime probed by PTAs, and potentially by future space-based detectors such as $\mu$Ares~\cite{Sesana:2019vho}. We derived the beyond-GR observables, especially the inter-pulsar cross correlation due to a single loud GW source. This framework, a direct extension of~\cite{mingarelli2026fingerprintsindividualsupermassiveblack}, provides an alternative to coherent matched filtering for analyzing potential beyond-GR effects in CW signals. We analyze the efficacy of this approach for three classes of beyond-GR modifications and demonstrate its practicality through a suite of injection-and-recovery tests. Complementing existing tests based on stochastic backgrounds~\cite{Cordes:2024oem, Bernardo:2023zna, Chamberlin:2011ev, Callister_2017, Yunes2025, Inomata:2024kzr, NANOGrav:2023ygs}, stellar-mass compact-binary mergers~\cite{LIGOScientific:2021sio, Takeda_2022, Schumacher:2025plq}, and indirect probes of gravitational potential~\cite{Desai_2018, loeb2024newlimitgravitonmass, PhysRevLett.123.161103}, this framework exploits the long-baseline phase coherence of PTAs and their unique nHz frequency coverage to provide an independent probe of gravitational physics and potential deviations from GR.

We have considered three classes of beyond-GR modifications: alternative polarizations, modified dispersion relations, and birefringence due to PV theories. Both alternative polarizations and modified dispersion relations are within the reach of PTAs. Birefringence effects, however, are highly suppressed in the ultra-low frequency range, and current observational limits on PV theories suggest they are not detectable by PTAs in practice; high-frequency experiments such as ground-based detectors are more suitable for such tests.

Three results stand out. First, for non-tensorial polarizations the CW cross correlation scales linearly in the alternative-polarization amplitude, compared to the quadratic scaling of the GWB, making CWs the more sensitive probe of additional polarization content once individual sources are detected. Second, the pulsar-term phase delay provides a distinct probe of modified dispersion relations, accessible to PTAs as pulsar-distance measurements improve (Fig.~\ref{fig:graviton_mass}). Third, a pure-GR CW template recovers source parameters without significant bias even when beyond-GR physics is present in the data, supporting a two-stage analysis strategy: identify candidates under GR, then test for deviations.

Although we have derived the cross correlations separately, a specific theory may contain a combination of these effects. For instance, a modified dispersion relation is usually accompanied by non-tensorial polarizations~\cite{FierzPauli1939, Mirshekari_2012}. One example is massive gravity, where the additional three degrees of freedom introduced by the graviton mass manifest as vector and scalar polarization modes. Other modified-gravity theories also predict superluminal propagation for the non-tensorial polarizations~\cite{Schumacher:2023jxq, Schumacher:2025plq}, which manifests as a modified dispersion relation for the alternative polarizations. Given current limits, the non-tensorial signals can precede the tensorial signals by years, making long-baseline observations in the nHz range especially useful for constraining such theories. A comprehensive search combining multiple classes of modifications self-consistently in the cross correlation could test multiple theories of gravity at once.

In this work, we have mostly focused on the Earth-term cross correlation, but the pulsar terms will also provide additional information inaccessible to the Earth term alone. The cross correlations may be extended to include the pulsar term rather straightforwardly, where a simplified case for modified dispersion relations is derived in Appendix~\ref{app:low frequency dispersion correlation}. Various physical effects could lead to a measurable phase shift in the pulsar term. First, the GW frequency evolution between Earth and the pulsar induces a measurable phase in the pulsar term, on top of the different frequencies during the observation time at the pulsar. The frequency evolution depends on waveform modeling, which is sensitive to binary properties such as mass ratio, eccentricity, spin-orbit coupling, and environmental effects~\cite{Schmidt:2014iyl, Cole:2022yzw}. Second, as we have shown in Sec.~\ref{sec:dispersion}, the modified dispersion relation directly induces a phase shift in the pulsar term. Including the pulsar term once precise pulsar distance measurements become available is a natural next step. It will be useful to systematically analyze the relative amplitudes of the phase shifts from these effects and map the correspondence between pulsar-distance measurement precision and the resulting constraining power. The analysis at the end of Sec.~\ref{sec:injection_and_recovery} provides a simplified example of this.

Several other directions remain for future investigation. First, although the framework developed here is formulated in the context of PTAs, it may also be applicable to future space-based detectors targeting ultra-low GW frequencies, such as $\mu$Ares~\cite{Sesana:2019vho}. In such configurations, each satellite pair could play a role analogous to the Earth–pulsar pair in PTAs. With sufficiently precise control and measurement of the satellite separations, effects analogous to the pulsar term may be measurable and incorporated into the analysis, potentially enabling similar tests of gravitational theories in a different observational setting. Multi-band tracking of the pulsar phase evolution may also provide unique opportunities for detecting deviations from GR, as discussed in~\cite{Zheng:inprep:MultibandGravityEchoes, Criswell:inprep:ArchivalMultiband}.

Higher-order-modes are
suppressed relative to the quadrupolar radiation by increasing powers of the orbital velocity $v/c$
(see Appendix~\ref{app:higher_modes} for explicit expressions, and~\cite{kidder_2008} for tabulated spin-weighted spherical harmonics), and are subleading to the injected beyond-GR signals in this work. However, as the search sensitivity improves, they should be incorporated in the analysis, together with the spin effects and binary eccentricity as discussed in~\cite{mingarelli2026fingerprintsindividualsupermassiveblack}. Such an extension will be straightforward following the existing derivation.

In addition, the injection-and-recovery demonstrations presented in this work consider CW signals embedded only in instrumental white noise. A more realistic noise environment---including intrinsic pulsar red noise and a GWB---is expected to affect the detection sensitivity and parameter recovery performance of the method. Extending the analysis to include these additional noise components and quantifying their impact will be an important step toward developing this work into a practical data-analysis pipeline.

Finally, applying the framework to real PTA datasets will be an important step in assessing its practical performance, testing the methodology under realistic observational conditions and providing a direct avenue for probing gravity theories using CW candidates identified in PTA data.

\section*{Software}
This work made use of \texttt{pta\_replicator}~\cite{Becsy:2025:pta_replicator} to simulate PTA datasets, \texttt{enterprise}~\cite{ENTERPRISE} to construct likelihoods and priors, and \texttt{nautilus}~\cite{nautilus} for nested sampling. We also used \texttt{NumPy}~\cite{harris_array_2020}, \texttt{SciPy}~\cite{2020SciPy-NMeth} and \texttt{matplotlib}~\cite{hunter_matplotlib_2007} for analysis and visualization.

\section*{Acknowledgments}
The authors would like to thank Kristen Schumacher Aloh and Michael Boyle for valuable feedback. C.M.F.M. thanks the Center for Computational Astrophysics (CCA) of the Flatiron Institute. The Flatiron Institute is supported by the Simons Foundation. C.M.F.M.\ was supported in part by the National Science Foundation under Grants No.\ NSF PHY-1748958, AST-2106552, and NASA LPS 80NSSC24K0440.

\appendix

\section{Cross correlation for Non–Tensorial Polarizations}
\label{app:polarizations}
In this appendix we document the explicit Earth–term correlation calculation to all six metric polarizations allowed in general metric theories of gravity, in the computational frame (see Fig.1 in~\cite{mingarelli2026fingerprintsindividualsupermassiveblack}). If multiple polarizations are present, there will be in general crossing terms involving different polarization modes, e.g., $F_a^+F_b^{\times},  F_a^+F_b^b$, etc. Instead we focus on what the cross correlation due to each individual polarization mode alone will look like, and ignore the cross coupling among modes. It is illustrative of why polarizations can be differentiated in a cross correlation framework.
For a single polarization state $P$, the Earth–term correlation is written as
\begin{equation}
\left\langle s_a(t)s_b(t)\right\rangle
=
\frac{A_{\rm CW}^2}{2}\,
\Upsilon^{(P)}_{ab}(\theta,\phi,\zeta).
\end{equation}
Here $\theta$, $\phi$ encodes the sky location of the source, while $\zeta$ is the angular separation of pulsar $a$ and $b$. Up to some normalization factor, the cross-correlation pattern is
\begin{equation}
\Upsilon^{(P)}_{ab}(\theta,\phi,\zeta)
=
F_a^P(\theta,\phi)\,
F_b^P(\theta,\phi,\zeta),
\label{eq:Uab-general-app}
\end{equation}
and $F_a^P$ is the Earth–term antenna pattern
\begin{equation}
F_a^P(\hat\Omega)
=
\frac{1}{2}\,
\frac{
\hat p_a^{\,i}\hat p_a^{\,j}\,
e^P_{ij}(\hat\Omega)
}{
1+\hat\Omega\cdot \hat p_a
}.
\label{eq:F-general}
\end{equation}

\subsection{Geometry of the computational frame}

In practical computations, we may define a convenient coordinate system as follows:
\begin{itemize}
    \item Pulsars:
\[
\hat p_a = (0,0,1),
\qquad
\hat p_b = (\sin\zeta,0,\cos\zeta).
\]
\item GW propagation direction:
\[
\hat\Omega
=
(\sin\theta\cos\phi,\
 \sin\theta\sin\phi,\
 \cos\theta).
\]

\item Transverse basis:
\begin{align}
\hat m &= (\sin\phi,\ -\cos\phi,\ 0), \\
\hat n &= (\cos\theta\cos\phi,\ \cos\theta\sin\phi,\ -\sin\theta).
\end{align}

\end{itemize}

\subsection{Polarization tensors}

\begin{align}
e^+_{ij} &= m_i m_j - n_i n_j, \\
e^\times_{ij} &= m_i n_j + n_i m_j, \\
e^{\rm b}_{ij} &= m_i m_j + n_i n_j, \\
e^{\rm L}_{ij} &= \sqrt{2}\,\Omega_i\Omega_j, \\
e^{\rm x}_{ij} &= m_i\Omega_j + \Omega_i m_j, \\
e^{\rm y}_{ij} &= n_i\Omega_j + \Omega_i n_j.
\end{align}

\subsection{Breathing mode}

Using Eq.~(\ref{eq:F-general}),

\[
F_a^{\rm b}(\theta)
=
\frac{1}{2}\,
\frac{\sin^2\theta}{1+\cos\theta},
\]

\[
F_b^{\rm b}(\theta,\phi,\zeta)
=
\frac{1}{2}
\frac{
\sin^2\phi\,\sin^2\zeta
+
\left(
-\sin\theta\cos\zeta
+
\sin\zeta\cos\phi\cos\theta
\right)^2
}{
1
+
\sin\theta\sin\zeta\cos\phi
+
\cos\theta\cos\zeta
}.
\]

Thus the breathing–mode correlation is

\begin{align}
&\Upsilon^{({\rm b})}_{ab}(\theta,\phi,\zeta)\notag\\
&=
\frac{\sin^2\theta}{4(1+\cos\theta)}
\,
\frac{
\sin^2\phi\,\sin^2\zeta
+
\left(
-\sin\theta\cos\zeta
+
\sin\zeta\cos\phi\cos\theta
\right)^2
}{
1
+
\sin\theta\sin\zeta\cos\phi
+
\cos\theta\cos\zeta
}.
\label{eq:U-b-full}
\end{align}

\subsection{Longitudinal scalar mode}

\[
F_a^{\rm L}(\theta)
=
\frac{\sqrt{2}}{2}
\frac{
\cos^2\theta
}{
1+\cos\theta
},
\]

\[
F_b^{\rm L}(\theta,\phi,\zeta)
=
\frac{\sqrt{2}}{2}
\frac{
\left(
\sin\theta\sin\zeta\cos\phi
+
\cos\theta\cos\zeta
\right)^2
}{
1
+
\sin\theta\sin\zeta\cos\phi
+
\cos\theta\cos\zeta
}.
\]

Therefore

\begin{equation}
\Upsilon^{({\rm L})}_{ab}(\theta,\phi,\zeta)
=
\frac{
\cos^2\theta
}{
2(1+\cos\theta)
}
\,
\frac{
\left(
\sin\theta\sin\zeta\cos\phi
+
\cos\theta\cos\zeta
\right)^2
}{
1
+
\sin\theta\sin\zeta\cos\phi
+
\cos\theta\cos\zeta
}.
\label{eq:U-L-full}
\end{equation}

\subsection{Vector-x mode}
For vector modes, the antenna pattern is transverse basis dependent. This means that 

\[
F_a^{\rm x} = 0,
\]

\[
F_b^{\rm x}(\theta,\phi,\zeta)
=
\frac{
\left(\sin\phi\,\sin\zeta\right)
\left(
\sin\theta\sin\zeta\cos\phi
+
\cos\theta\cos\zeta
\right)
}{
1
+
\sin\theta\sin\zeta\cos\phi
+
\cos\theta\cos\zeta
}.
\]

Thus in this computational frame

\begin{equation}
\Upsilon^{({\rm x})}_{ab}(\theta,\phi,\zeta)
= 0.
\label{eq:U-x-full}
\end{equation}

\subsection{Vector-y mode}

\begin{equation}
F_a^{\rm y}(\theta)
=
-\frac{\sin(2\theta)}{2(1+\cos\theta)},
\end{equation}

\begin{align}
&F_b^{\rm y}(\theta,\phi,\zeta)\notag\\
&=
\frac{
\left(
-\sin\theta\cos\zeta
+
\sin\zeta\cos\phi\cos\theta
\right)
\left(
\sin\theta\sin\zeta\cos\phi
+
\cos\theta\cos\zeta
\right)
}{
1
+
\sin\theta\sin\zeta\cos\phi
+
\cos\theta\cos\zeta
}.
\end{align}

Thus
\begin{widetext}
\begin{equation}
\Upsilon^{({\rm y})}_{ab}(\theta,\phi,\zeta)
=
-\frac{\sin(2\theta)}{2(1+\cos\theta)}
\,
\frac{
\left(
-\sin\theta\cos\zeta
+
\sin\zeta\cos\phi\cos\theta
\right)
\left(
\sin\theta\sin\zeta\cos\phi
+
\cos\theta\cos\zeta
\right)
}{
1
+
\sin\theta\sin\zeta\cos\phi
+
\cos\theta\cos\zeta
}.
\label{eq:U-y-full}
\end{equation}
\end{widetext}

\section{Basis transformations of antenna patterns}
\label{app:basis_transformations}
The explicit forms of the antenna pattern functions depend on the transverse basis. Throughout the paper we have used antenna patterns defined in both the basis where $\psi$ is explicitly zero (e.g., in writing down the waveforms in Eq.~(\ref{eq:SMBHB waveforms})) and the cosmic rest frame. The transformations between the two conventions introduce dependence on the polarization angle $\psi$, which in turn enters through the transverse basis transformation
\begin{align}
    \begin{pmatrix}
        \hat{m}' \\
        \hat{n}'
    \end{pmatrix} = \begin{pmatrix}
        \cos\psi & \sin\psi \\
        -\sin\psi & \cos\psi
    \end{pmatrix}\begin{pmatrix}
        \hat{m} \\
        \hat{n}
    \end{pmatrix}.
\end{align}
By the definitions of the antenna pattern functions in Sec.~\ref{sec:non-tensor}, the scalar transverse (breathing) mode and scalar longitudinal mode are basis independent. The vector modes transform as
\begin{align}
    \begin{pmatrix}
        F^x \\
        F^y
    \end{pmatrix}_{\psi=0} = \begin{pmatrix}
        \cos\psi & \sin\psi \\
        -\sin\psi & \cos\psi
    \end{pmatrix}\begin{pmatrix}
        F^x \\
        F^y
    \end{pmatrix}_{\rm crf}.
\end{align}
The tensor modes transform as
\begin{align}
\label{eq:antenna_patterns}
    \begin{pmatrix}
        F^+ \\
        F^{\times}
    \end{pmatrix}_{\psi=0} = \begin{pmatrix}
        \cos2\psi & \sin2\psi \\
        -\sin2\psi & \cos2\psi
    \end{pmatrix}\begin{pmatrix}
        F^+ \\
        F^{\times}
    \end{pmatrix}_{\rm crf}.
\end{align}
We choose to work in a specific basis for convenience in writing down the waveform, but the final result should be transformed to a well defined basis such as in the cosmic rest frame (crf).

\section{Derivations of birefringence cross-correlations and limits}
\label{app:birefringence}
In this appendix we derive the single source cross correlation in the presence of birefringence.

We start from the GR response at Earth defined in Eq.~(\ref{eq:GR_decomp}). In frequency domain the cross and plus polarizations become
\begin{align}
    h_+(\omega)=a\Big[e^{i\phi_0}\delta(\omega-\omega_0)\Big],\\
    h_{\times}(\omega)=ib\Big[e^{i\phi_0}\delta(\omega-\omega_0)\Big],
\end{align}
where $a=h_0(1+\cos^2\iota)\pi$ and $b=2h_0\cos\iota \pi$. Plugging into Eq.~(\ref{eq:GR_decomp}) one gets
\begin{align}
    h_a^{GR}(\omega)=(F_a^{+}a+iF_a^{\times}b)\Big[e^{i\phi_0}\delta(\omega-\omega_0)\Big].
\end{align}
Rewriting Eq.~(\ref{eq:h_af}) as 
\begin{equation}
    h_a(\omega)=h_a^{GR}(\omega)\\\varphi_a(\omega,\vec{\theta}),
\end{equation}
with
\begin{align}
\label{eq:varphidef}
    \varphi_a(\omega,\vec{\theta})= \left[
1
+ f(F_a^{+, \times}, \xi)\,\delta\phi_{A}(\omega)
- g(F_a^{+, \times}, \xi)\,\delta\phi_{V}(\omega)
\right]\notag\\
\exp\!\left\{
i \left[
g(F_a^{+, \times}, \xi)\,\delta\phi_{A}(\omega)
+ f(F_a^{+, \times}, \xi)\,\delta\phi_{V}(\omega)
\right]
\right\},
\end{align}
where $\vec{\theta}=\{\alpha,\beta,\rho, \sigma, z, \hat{\Omega}, \hat{p}_a,\xi, \psi\}$ contains the source and pulsar locations as well as the source inclination, on top of the parameters characterizing the birefringence.

It the follows from Eq.~(\ref{eq:s_at}) that
\begin{align}
\label{eq:saresult}
    s_a(t)&=\int^tdt'\int d\omega\, h_a(\omega)e^{i\omega t'}\notag\\
    &=\int d\omega \, h_a(\omega)\int^t dt'\, e^{i\omega t'}\notag\\
    &=\int d\omega\, h_a(\omega) \frac{1}{i\omega}e^{i\omega t}\notag\\
    &=(F_a^{+}a+iF_a^{\times}b)\frac{1}{i\omega_0}\cdot
    \Big[\varphi_a(\omega_0, \vec{\theta})e^{i(\omega_0 t+\phi_0)}\Big]
\end{align}
Since the physical timing residual is real, taking the real part of the result in Eq.~(\ref{eq:saresult}) gives the general form
\begin{equation}
    s_a(t)=A_a(\omega_0,\vec{\theta})\cos(\omega_0t+\phi_0)+B_a(\omega_0, \vec{\theta})\sin(\omega_0t+\phi_0),
\end{equation}
or alternatively
\begin{equation}
    s_a(t)=A^a_{CW}(\omega_0, \vec{\theta})\sin(\omega_0t+\phi_0+\delta_a(\omega_0, \vec{\theta})).
\end{equation}
Here we have introduced a phase $\delta_a(\omega_0, \vec{\theta})=\arctan(\frac{A_a}{B_a})$, and $A^a_{CW}=\frac{A_a}{\sin\delta_a}=\frac{B_a}{\cos\delta_a}$. Finally the cross correlation can be calculated by taking the time average
\begin{align}
    &\langle s_a(t)s_b(t)\rangle_T\notag\\
    &=\frac{A^a_{CW}A^b_{CW}}{T}\int_0^T dt \, \sin(\omega_0t+\phi_0+\delta_a)\sin(\omega_0t+\phi_0+\delta_b)\notag\\
    &=\frac{A^a_{CW}A^b_{CW}}{2}\cos(\delta_a-\delta_b)\notag\\
    &-\frac{A^a_{CW}A^b_{CW}}{4\omega_0T}\Big[\sin(2\omega_0 T+2\phi_0+\delta_a+\delta_b)-\sin(2\phi_0+\delta_a+\delta_b)\Big]\notag\\
    &\simeq \frac{A^a_{CW}A^b_{CW}}{2}\cos(\delta_a-\delta_b).
\end{align}
In the last step we used the assumption that the source is non-evolving, hence the second term is a boundary term suppressed by $1/(\omega T)$. Rewriting Eq.~(\ref{eq:varphidef}) in a compact form as $\varphi_a(\omega,\vec{\theta})=p_a(\omega,\vec{\theta})+iq_a(\omega,\vec{\theta})$ where $p_a, q_a$ are real, we may solve for $C^{a,\, b} _{CW}$ and $\delta_{a,\, b}$ explicitly as 
\begin{align}
\label{eq:delta_and_A}
    &\delta_{a,\, b}=\arctan\left(\frac{b\,  p_{a,b} F_{a,\, b}^\times+a\, q_{a,b} F_{a,\, b}^+}{a\, p_{a,b} F_{a,\, b}^+ -b\, q_{a,b} F_{a,\, b}^\times}\right),\notag\\
    &A^{a,\, b} _{CW}=\frac{1}{\omega}\frac{b\, p_{a,b} F_{a,\, b}^\times+a\, q_{a,b} F_{a,\, b}^+}{\sin \delta_{a,\, b}}.
\end{align}
The antenna patterns are defined in Eq.~(\ref{eq:antenna_patterns}), with $\psi$ dependence. Here we assume that the birefringence modification is small, such that locally the dispersion can be ignored $\omega\approx ck$, and $k$'s in Eq.~(\ref{eq:PV_parameterization}) can be substituted with $\omega$. This is consistent with the idea that birefringence effects are measurable through accumulated deviations over long propagation distances, rather than local dispersion. In principle, the dispersion also modifies the PTA antenna patterns. Specifically, the denominator becomes chirality dependent, with 
\begin{equation}
    v_p^{R, L}=1+\frac{\lambda_{R, L}}{2}\Big(\frac{\rho \mathcal{H}}{k}+\frac{\sigma k}{a}\Big),
\end{equation}
where $\lambda_{R, L}=+\,,\times$. We ignore this chiral dependence for now using again that the deviations from GW are small. This is justified by the current limits on the expansion parameters. In the local Universe, for $f_{\rm gw}\sim 10\, \rm nHz$, $\rho \mathcal{H}/k\sim \mathcal{O}(10^{-10})$, and $\sigma \,k/a\sim \mathcal{O}(10^{-33})$. 

To estimate the birefringence induced pulsar term phase shift, we recall Eq.~(\ref{eq:dispersion_phase}), and rewrite it as
\begin{equation}
    \Delta\Phi_a^{\rm Disp}=-\omega d_a\Big(\frac{1}{v_p^{R, L}}-1\Big)\hat{\Omega}\cdot \hat{p}_a.
\end{equation}
We may then estimate the chiral imbalance between the right and left handed circular polarizations in the phase shift as
\begin{equation}
    \delta\Delta\Phi_a^{\rm Bire}=-\omega d_a\Big(\frac{\rho \mathcal{H}}{\omega}+\frac{\sigma \omega}{a}\Big)\hat{\Omega}\cdot \hat{p}_a.
\end{equation}
With $d_a\sim 1\, \rm kpc$ and $\omega\sim 10\, \rm nHz$, the phase shift $\delta\Delta\Phi_a^{\rm Bire}\sim \mathcal{O}(10^{-8})$. As a result, the birefringence modifications to neither the antenna patterns nor the pulsar term phase shift are significant in practice. Finally, as a consistency check, in the limit where $\delta \phi_A=\delta \phi_V =0$, the cross correlation recovers the GR case.

To give an order-of-magnitude estimate of the waveform modifications due to birefringence, as presented in Fig.~\ref{fig:birefringence}, we define the parameter $\delta\phi$ in Eq.~(\ref{eq:deltaphi}). We then compute the current constraints on $\alpha$, $\beta$, $\rho$, and $\sigma$ using data in~\cite{Jenks:2023pmk, Zhao_2022, PhysRevD.108.084068}. Assuming $z\sim\mathcal{O}(10^0)$ and with constraint in~\cite{Zhao_2022}, we estimate $\alpha + \beta \,d_2(z)\sim 10^{-2} \, \rm Hz^{-1}$; with constraint from~\cite{PhysRevD.108.084068}, we estimate $\sigma \, d_3(z)\sim 10^{-8} \, \rm Hz^{-2}$. The parameter $\rho$ induces a frequency-independent overall phase shift of the signal, and from~\cite{Wang:2025fhw} we estimate its limit to be $\rho\sim\mathcal{O}(10^1)$. 

\begin{figure*}[ht!]
    \centering
    \includegraphics[width=0.49\linewidth]{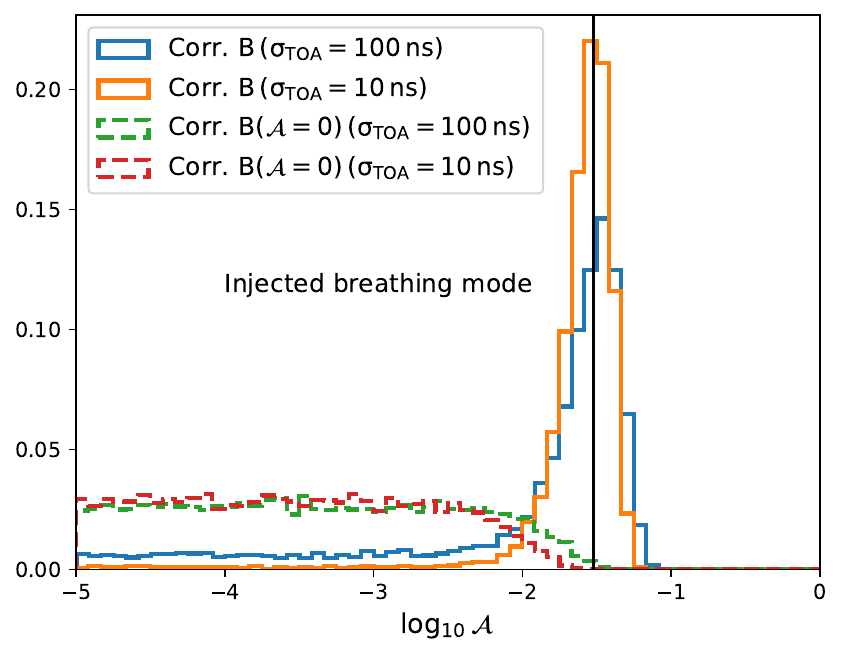} 
    \includegraphics[width=0.493\linewidth]{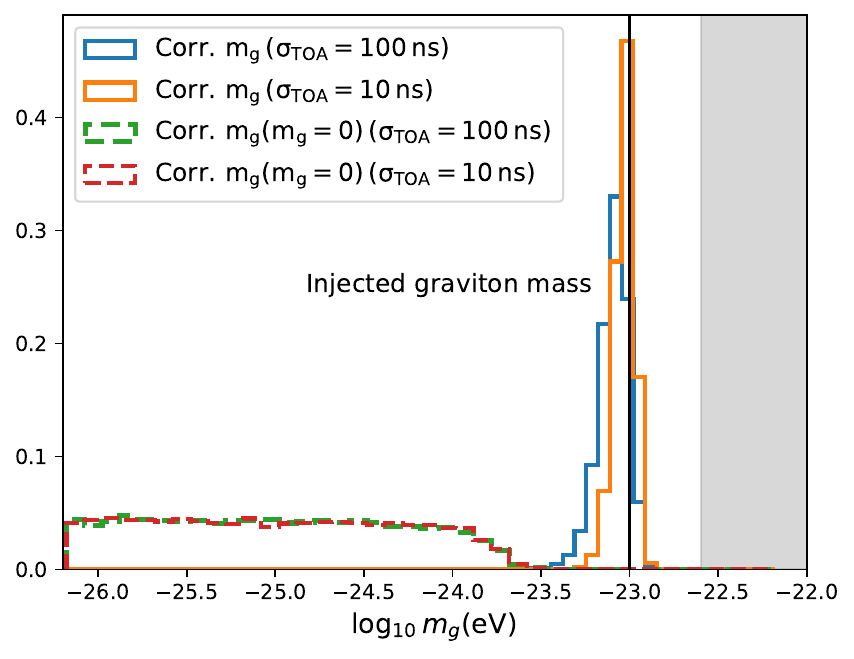}
    \caption{An improving timing precision results in better recovery of the injected beyond-GR parameters with higher Bayes factors. Top: the posteriors of $\mathcal{A}$ between the runs with $\sigma_{\rm TOA}=100\, \rm ns$ and $\sigma_{\rm TOA}=10\, \rm ns$, both with the cross-correlated CW model. The $\sigma_{\rm TOA}=10\, \rm ns$ posterior has a higher peak at the injected value and a higher Bayes factor of $\log\mathcal{B}_B=3.0$. The better timing precision also provides a tighter constraint on the breathing mode amplitude when there is no injected breathing mode. Bottom: the posteriors of $m_g$ between the runs with $\sigma_{\rm TOA}=100\, \rm ns$ and $\sigma_{\rm TOA}=10\, \rm ns$, both with the cross-correlated CW model. The $\sigma_{\rm TOA}=10\, \rm ns$ posterior also has a higher peak at the injected value and a higher Bayes factor of $\log\mathcal{B}_{m_g}=10.5$, as in the breathing mode case. However, both timing precision runs put similar constraints on the graviton mass. The shaded region corresponds to GW attenuation, where $\chi$ becomes imaginary (Eq.~\ref{eq:massive_graviton_antenna_patterns}).}
\label{fig:beyond_GR_100ns_10ns}
\end{figure*}
\section{Low frequency limit cross correlation with modified dispersion relations}
\label{app:low frequency dispersion correlation}
We additionally derive the cross correlation with the modified dispersion relations, in the low frequency limit where the GW frequency evolution may be neglected relative to the expected modification to the dispersion. For example, referring to Fig.~\ref{fig:graviton_mass}, when the graviton mass is $10^{-23}\, \rm eV$ for a $1\, \rm nHz$ GW emitted by a $10^7\, M_{\odot}$ SMBHB, the limit considered here is justified. We start from the full expression of the timing residual
\begin{align}
    s_+(t) &= \frac{h_0}{\omega_0} (1+\cos^2\iota)\Bigg[\sin\Big(\omega_0 t + \phi_0\Big)\notag\\
    &-\sin\Big(\omega_0 t + \phi_0-\omega_0 d_a(1+\frac{1}{v_p}\hat \Omega\cdot \hat{p}_a)\Big)\Bigg],\notag\\
    s_\times(t) &= \frac{2 h_0}{\omega_0} \cos\iota\Bigg[ \cos\Big(\omega_0 t + \phi_0\Big)\notag\\
    &-\cos\Big(\omega_0 t + \phi_0-\omega_0 d_a(1+\frac{1}{v_p}\hat \Omega\cdot \hat{p}_a)\Big)\Bigg].
\end{align}
Then the cross correlation of the full timing residual is 
\begin{align}
    C_{ab}&=\left<s_a(t)s_b(t)\right>\notag\\
    &=\frac{A_{CW}^2}{2}\Bigg[A_aA_b\cos(\delta_a-\delta_b)+A_a\gamma_{1b}\cos(\delta_a-\varphi_{1b})\notag\\
    &+A_b\gamma_{1a}\cos(\delta_b-\varphi_{1a})+A_a\gamma_{2b}\cos(\delta_a-\varphi_{2b})\notag\\
    &+A_b\gamma_{2a}\cos(\delta_b-\varphi_{2a})+\gamma_{1a}\gamma_{2b}\cos(\varphi_{1a}-\varphi_{2b})\notag\\
    &+\gamma_{1b}\gamma_{2a}\cos(\varphi_{1b}-\varphi_{2a})+\gamma_{1a}\gamma_{1b}\cos(\varphi_{1a}-\varphi_{1b})\notag\\
    &+\gamma_{2a}\gamma_{2b}\cos(\varphi_{2a}-\varphi_{2b})\Bigg],
\end{align}
where $A_{a,b}, A_{CW}, \delta_{a, b}$ follow the definition in the previous sections and 
\begin{align}
    \gamma_{1a}&=-(1+\cos^2\iota)F_+^a\notag\\
    \gamma_{2a}&=-2\cos\iota F_{\times}^a\notag\\
    \varphi_{1a}&=-\omega_0 d_a(1+\frac{1}{v_p}\hat \Omega\cdot \hat{p}_a)\notag\\
    \varphi_{2a}&=-\omega_0 d_a(1+\frac{1}{v_p}\hat \Omega\cdot \hat{p}_a)+\frac{\pi}{2}
\end{align}

\section{Injection-and-recovery tests}
\subsection{Bayes factors}
\label{app:Bayes_factors}
The Bayes factors used in the model comparisons, which we denote with $\mathcal{B}_{\rm Model\, 2}^{\rm Model\, 1}$, are calculated with
\begin{equation}
    \log \mathcal{B}_{\rm Model\, 2}^{\rm Model\, 1}=\log Z_{\rm Model\, 1} - \log Z_{\rm Model\, 2},
\end{equation}
where $Z_{\rm Model\, 1}$, $Z_{\rm Model\, 2}$ are the evidence for the two models.

For the particular case where the two models are the cross-correlated CW models with and without the beyond-GR modification, the Bayes factor can also be computed by the Savage Dickey approach where the GR model is taken as the continuous limit for the beyond-GR model. This Bayes factor is defined as the Savage-Dickey density ratio
\begin{equation}
    \mathcal{B}^{\rm beyond-GR}_{\rm GR}=\frac{p(\theta=\theta_0 | M_{CC})}{p(\theta=\theta_0 | d, M_{CC})},
\end{equation}
where $M_{CC}$ is the beyond-GR cross-correlated model, $\theta$ are the generic parameters controlling the beyond-GR effects, and $\theta_0$ recovers the GR limit. We have checked that the Bayes factors computed from the evidence method and the Savage Dickey method are consistent.

\subsection{Beyond-GR parameter recovery with improving SNRs}
\label{app: recovery_SNRs}
We show that the beyond-GR parameter recovery may be improved with higher SNRs. One way to achieve a higher SNR is through more precise timing. In the main injection-and-recovery test in Section~\ref{sec:injection_and_recovery}, we take the timing precision to be $\sigma_{\rm TOA}=100\, \rm ns$. Here we adopt $\sigma_{\rm TOA}=10\, \rm ns$, with other injected parameters held the same. The SNR scales inversely with $\sigma_{\rm TOA}$, and indeed the SNR for the $\sigma_{\rm TOA}$ data turns out to be around $\rm SNR\approx 640$. 

For both breathing mode and massive gravity, the beyond-GR parameters have a more concentrated posterior distribution around the injected value when the SNR is higher (Fig.~\ref{fig:beyond_GR_100ns_10ns}). For breathing mode, the improved Savage-Dickey Bayes factor sits at $\log \mathcal{B}_B=3.0$; for massive gravity, the improved Bayes factor is $\log\mathcal{B}_{m_g}=10.5$. More discussions about the interpretation of the Bayes factors can be found in Section~\ref{sec:injection_and_recovery}.

\subsection{Joint posteriors of injection-and-recovery tests}
\label{app: full_posteriors}
\begin{figure*}

    \centering
    \includegraphics[width=1\linewidth]{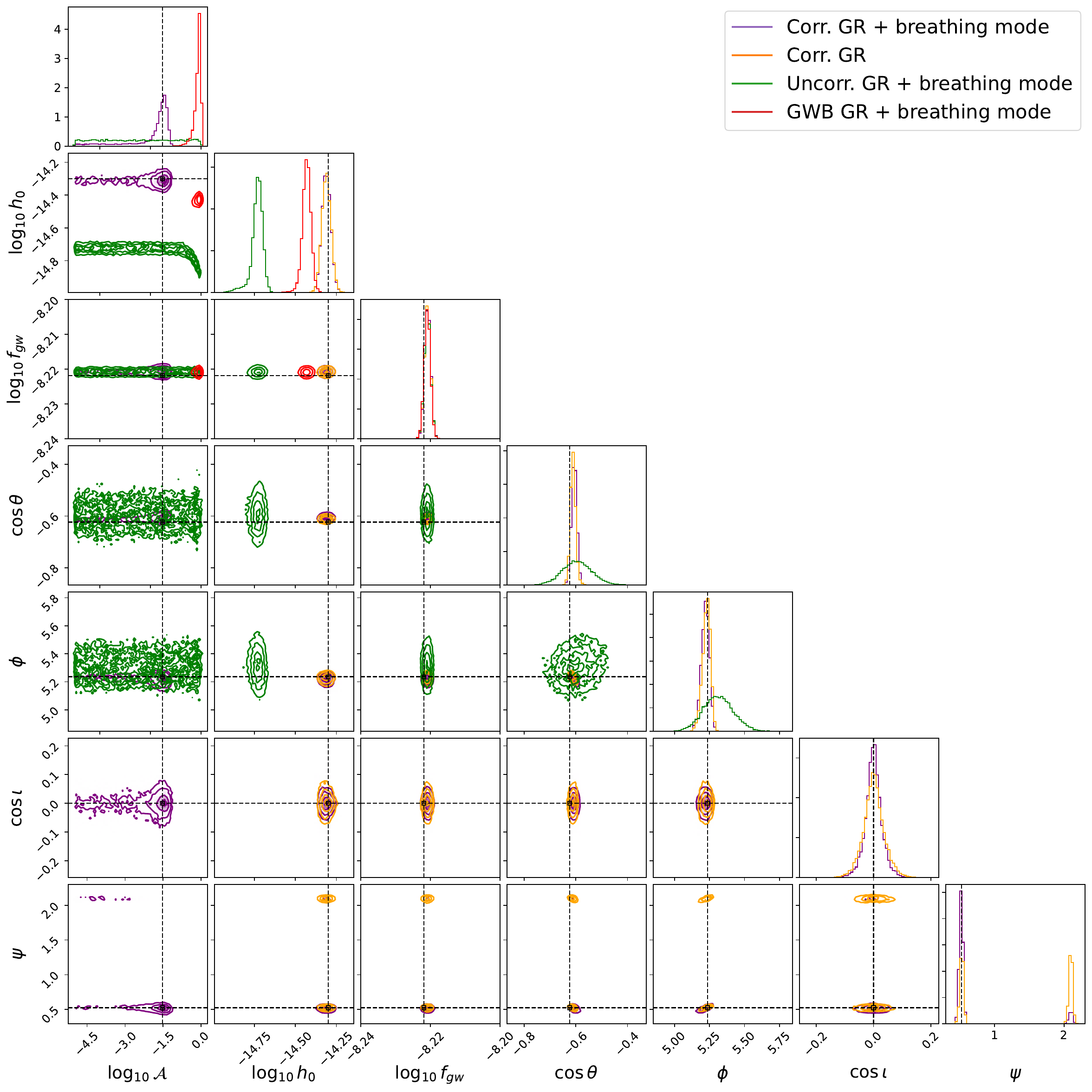}
    \caption{Joint posterior distributions for data containing a breathing mode in addition to the transverse tensor modes, with the injected values indicated by black dashed lines. Four models are used in the analysis: (1) a cross-correlated CW model including both the GR tensor modes and a breathing mode; (2) a cross-correlated CW model including only the GR tensor modes; (3) an uncorrelated CW model including the GR tensor modes and a breathing mode; and (4) a cross-correlated GWB model including the GR tensor modes and a breathing mode. Only model (1) faithfully recovers the injected breathing mode. Model (2) performs similarly to model (1) in recovering the other parameters. Model (3) yields less concentrated posterior distributions than models (1) and (2), consistent with the findings of Ref.~\cite{mingarelli2026fingerprintsindividualsupermassiveblack}. Both models (3) and (4) also introduce a bias in the recovered strain amplitude of the signal.}
    \label{fig:full_posterior_breathing_mode}
\end{figure*}

We include the joint posterior distributions of the searches on data with beyond-GR signals injected. In Fig.~\ref{fig:full_posterior_breathing_mode}, we overlay the search results on the CW data with an additional breathing mode injected using four different models introduced in Section~\ref{sec:injection_and_recovery}. Notice that the introduction of the breathing mode to the model breaks the degeneracy of the polarization angle $\psi$.

\begin{figure*}

    \centering
    \includegraphics[width=1\linewidth]{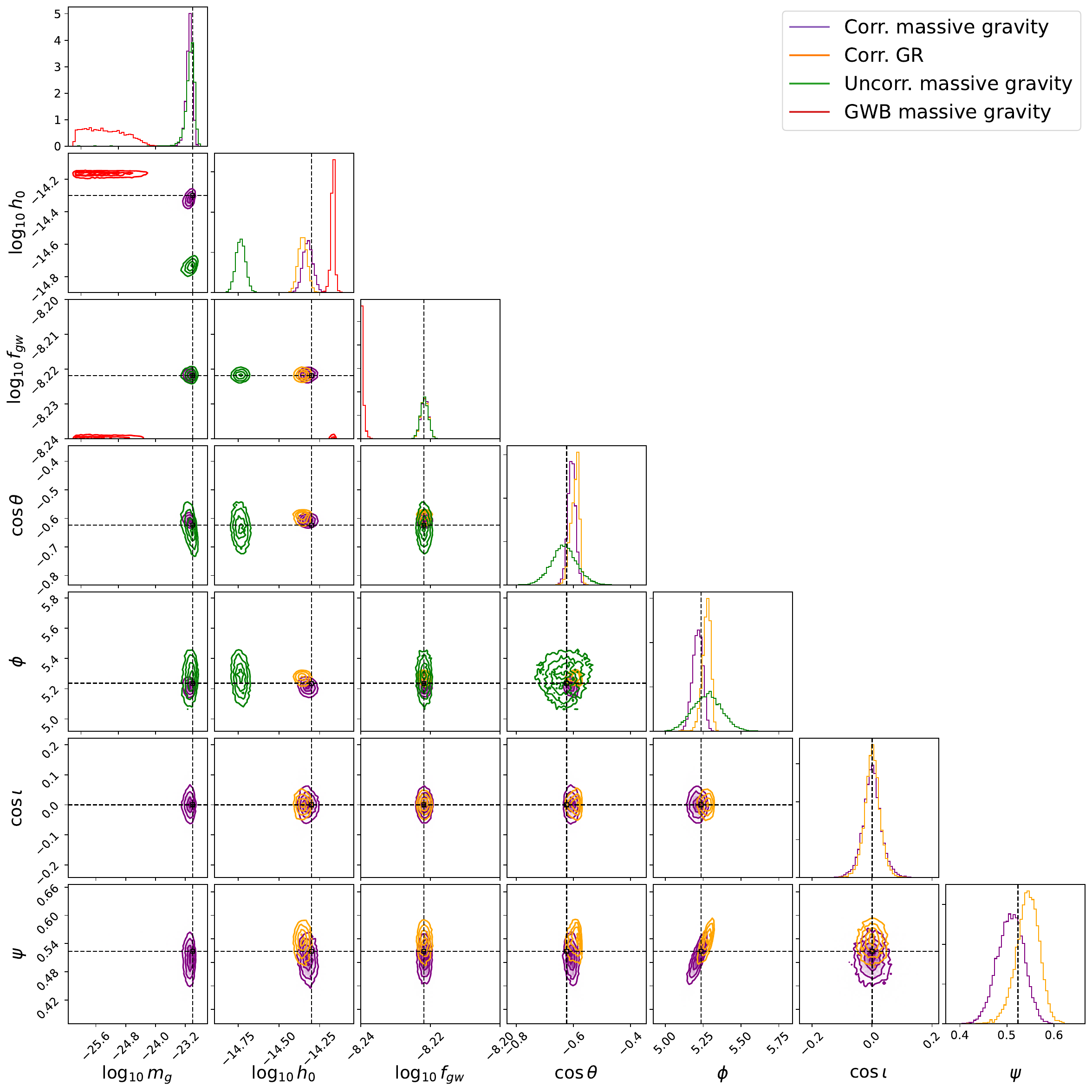}
    \caption{Joint posterior distributions for data containing a massive graviton signal, with the injected values indicated by black dashed lines. Four models are used in the analysis: (1) a cross-correlated CW model allowing for a massive graviton; (2) a cross-correlated CW model assuming GR; (3) an uncorrelated CW model allowing for a massive graviton; and (4) a cross-correlated GWB model allowing for a massive graviton. Both models (1) and (3) recover the injected graviton mass. However, model (3) yields less concentrated posterior distributions than model (1) for the source sky location. Model (2) recovers the CW parameters comparably well to model (1), except of course for the graviton mass. Model (4) fails to recover the graviton mass and also recovers the CW frequency significantly more poorly than the other models, with a clear offset from the injected value. In addition, both models (3) and (4) introduce a bias in the recovered strain amplitude.}
    \label{fig:full_posterior_massive_graviton}
\end{figure*}

In Fig.~\ref{fig:full_posterior_massive_graviton}, the search results are shown for four different models on the data with a CW signal in massive gravity. The discussion of these models are in Section~\ref{sec:injection_and_recovery}.

We emphasize that in both cases, the pure-GR cross-correlated model is able to reproduce the source parameters without severe bias induced by the beyond-GR physics in the data. As a result, searching for the CW sources with a pure-GR model and later including the beyond-GR effects is viable.

\section{Higher-order-mode corrections}
\label{app:higher_modes}
\begin{figure*}

    \centering
    \includegraphics[width=1\linewidth]{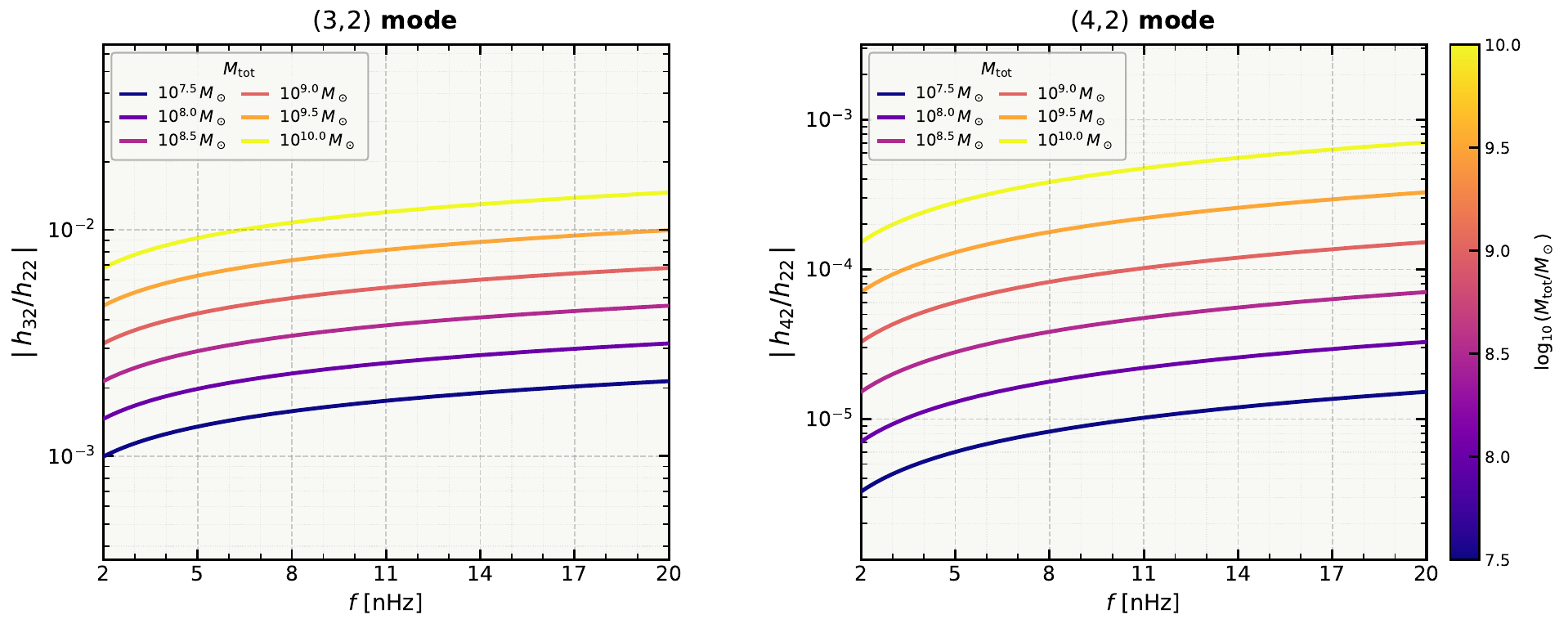}
    \caption{The higher-order modes are sub-dominant given the injected beyond-GR signals in the current analyses, but may eventually be important to include in the parametrized cross correlation. Here we show the relative amplitude of the $(3,2)$ mode and $(4,2)$ mode, with respect to the dominating quadrupolar mode. Lower frequencies and lower mass further suppress the PN corrections.}
    \label{fig: higher_modes}
\end{figure*}

The higher-order-mode corrections will become important when the search sensitivity improves. In the ideal case of a non-spinning quasi-circular SMBHB, the higher-order modes may still impact the beyond-GR analyses. The GW strain from a binary is decomposed using spin-weighted
spherical harmonics of weight $-2$,
\begin{equation}
    h_+ - i h_\times = \sum_{\ell \geq 2} \sum_{m=-\ell}^{\ell}
    h_{\ell m}(t) \; {}_{-2}Y_{\ell m}(\iota, \varphi_0),
\end{equation}
where $\iota$ is the inclination of the orbital angular
momentum vector with respect to the line of sight, and $\varphi_0$ is a reference azimuthal
angle. Within the PN formalism for quasi-circular orbits, the mode
amplitudes are power series in the dimensionless orbital velocity parameter
\begin{equation}
    \frac{v}{c} = \left(\frac{\pi G M f_{\mathrm{GW}}}{c^3}\right)^{1/3},
\end{equation}
where $M$ is the total binary mass and $f_{\mathrm{GW}}$ is the gravitational wave
frequency. Each mode $(\ell, m)$ radiates at frequency $|m|\,f_{\mathrm{orb}}$, so
restricting to $|m| = 2$ selects the subset of modes that emit at exactly
$f_{\mathrm{GW}} = 2\,f_{\mathrm{orb}}$, the same frequency as the dominant $(2,2)$ mode.
Setting $x \equiv (v/c)^2$, the leading-order amplitudes of the $|m| = 2$ modes are
\begin{equation}
    |h_{22}| \propto \eta\, x,
    \qquad
    |h_{32}| \propto \eta\,(1 - 3\eta)\, x^2,
    \qquad
    |h_{42}| \propto \eta\, x^3,
\end{equation}
where $\eta = m_1 m_2/M^2$ is the symmetric mass ratio~\cite{Arun:2008kb,Blanchet:2013haa}.
The amplitude ratio of the leading sub-dominant mode to the quadrupole is therefore
\begin{equation}
    \left|\frac{h_{32}}{h_{22}}\right| \approx C\,(1 - 3\eta)\,\left(\frac{v}{c}\right)^2,
    \label{eq:mode_ratio}
\end{equation}
where $C \approx O(1)$ is a numerical coefficient fixed by the PN expansion. The factor
$(1 - 3\eta)$ ranges from $1/4$ for equal-mass binaries ($\eta = 1/4$) to unity in the
extreme mass-ratio limit ($\eta \to 0$), so the suppression is dominated by the velocity
factor $(v/c)^2$ with at most a factor-of-four variation across mass ratios.

For a representative SMBHB with total mass $M = 10^9\,M_\odot$ at
$f_{\mathrm{GW}} = 10\,\mathrm{nHz}$, 
\begin{equation}
    \frac{v}{c}
    = \left(\pi \times GM/c^3
      \times f_{\rm GW}\right)^{1/3}
    \approx 0.054,
\end{equation}
giving $(v/c)^2 \approx 2.9 \times 10^{-3}$. Substituting into Eq.~\eqref{eq:mode_ratio}
and bounding $(1-3\eta) \leq 1$ gives $|h_{32}/h_{22}| \lesssim \mathcal{O}(10^{-3})$
across the full range of mass ratios. This is the highest-frequency end of the nHz band, where
the correction is largest; at lower frequencies the suppression is stronger. The
$(4,2)$ mode is further suppressed by an additional factor of $(v/c)^2 \approx 3 \times
10^{-3}$. Fig.~\ref{fig: higher_modes} shows the relative amplitude of these PN corrections for multiple binary total masses across the PTA band, suggesting the necessity of including such contributions as search sensitivity improves or when the target beyond-GR effects are small.

\bibliographystyle{apsrev4-2}
\bibliography{bib}

\end{document}